\documentclass[12pt]{iopart}
 
\usepackage{amsfonts}
\usepackage{amssymb} 
\usepackage{bm}
\usepackage{graphicx}
\usepackage[outdir=./]{epstopdf}
\usepackage[T1]{fontenc}

\begin{document}
 
\title[Interfaces as indicators of mutational instability]{Shape of  population interfaces as an indicator of mutational instability in coexisting cell populations}
\author{Clarisa Castillo}
\address{Department of Physics \& Astronomy, University of Tennessee, Knoxville, Tennessee 37996, USA}
\ead{clarisaecastillo@gmail.com}
\author{Maxim O. Lavrentovich}
\address{Department of Physics \& Astronomy, University of Tennessee, Knoxville, Tennessee 37996, USA}
\ead{lavrentm@gmail.com}

\begin{abstract}
Cellular populations such as avascular tumors and microbial biofilms may ``invade'' or grow into surrounding populations. The invading population is often comprised of a heterogeneous mixture of cells with varying growth rates. The population may also exhibit mutational instabilities, such as a heavy deleterious mutation load in a cancerous growth. We study the  dynamics of a heterogeneous, mutating population competing with a surrounding homogeneous population, as one might find in a cancerous invasion of healthy tissue. We find that the shape of the population interface  serves as an indicator for the evolutionary dynamics \textit{within} the heterogeneous population. In particular, invasion front undulations become enhanced when the invading population is near a mutational meltdown transition or when the surrounding ``bystander'' population is barely able to reinvade the mutating population.    We characterize these interface undulations  and the effective fitness of the heterogeneous population in one- and two-dimensional systems.    
\end{abstract}
\submitto{\PB}
\maketitle

\section{Introduction}

 Invasion and competitive exclusion is a common phenomenon in biology, with examples spanning a wide range of length  and time scales:  An invasive land animal species may compete with the species already present in the ecological habitat \cite{gecko}, microbial strains may compete and invade each other within a growing biofilm \cite{biofilm,biofilm2}, or virus strains may compete for host resources \cite{viruscompetition}. Such competitions also exist \textit{within} the tissues of various organisms, during development and in cancerous growth: a tumor which starts out as a small cluster of rapidly growing and mutating cells must compete with surrounding healthy tissue \cite{competitionreview}.   In all of these examples, the spatial structure of the population may have a significant impact on the strain competition and evolution.

Spatially-distributed populations are markedly different from their well-mixed counterparts. Because local population sizes are small compared to the population size in a well-mixed test tube, \textit{genetic drift} or small number fluctuations become more important: Strains within spatially-distributed populations are more likely to locally fix.   Also, deleterious mutations  more readily accumulate at leading edges of growing populations compared to well-mixed populations where natural selection would eliminate such variants \cite{delrange,ExcoffierDeleterious1}. There may be mitigating factors that reduce this mutational load, however, such as the presence of an Allee effect due to strain cooperation, for example \cite{FoutelRodierEtheridge}. These considerations are particularly important for invading cancerous populations which exhibit genomic instability \cite{tumorunstable,InstabilityReview} and are typically spatially heterogeneous, consisting of a wide distribution of strains \cite{tumorhet1,tumorhet2,tumorhet3,bigbang}. It is becoming increasingly clear that spatial evolutionary models are necessary to understand  the evolutionary dynamics of cancer cell populations \cite{rangetumor,korolev1}.

 The  mutations that drive uncontrolled growth in cancerous populations are  the so-called \textit{driver mutations}. However, the  majority of mutations are \textit{passenger mutations} which have a neutral or slightly deleterious effect on the cancer cells.   Such mutations are ubiquitous in cancerous populations, although their importance for cancer progression has only recently been recognized \cite{KorolevPNAS}. Weakly deleterious passenger mutations can rapidly accumulate at the edges of spatially-distributed populations, and the combined deleterious effect can lead to a cancer population collapse.  Therefore,  the elucidation of the impact of the passenger mutations may lead to new cancer therapies and a  better understanding of the efficacy of existing therapies  \cite{mcfarland1,mcfarland2}. Indeed, an effective cancer treatment may involve increasing the mutation rate such that the passengers overwhelm the drivers or increasing the deleterious effect of the passengers such that the drivers are no longer able to sustain  tumor growth.  The accumulation of deleterious mutations leading to population collapse is termed ``mutational meltdown'' \cite{muller,meltdown}.  Already there is evidence that cancer therapies may be developed that target  passenger mutations to expose vulnerabilities in  cancer growth \cite{therapy} and that cancer cell lines are particularly vulnerable to mutational meltdown \cite{cancermeltdown}.

In this paper, we develop     a simple spatial model of the invasion of a cellular population (i.e., a ``bystander'') by another population (i.e., an ``invader'') that is acquiring deleterious  mutations. We show that when the mutating invader is near a mutational meltdown, the interface between the invader and bystander becomes rougher and more undulated. Such interface shapes and physical cues are important as advances in medical imaging allow us to probe the spatial structure of cancerous growth with unprecedented detail \cite{cancerimage}. Tumor shape is increasingly being used for diagnostic purposes. For example, the shape of a tumor boundary is used as a diagnostic tool in breast cancers where a rougher tumor edge may indicate a more malignant growth \cite{breastimage}. Spatial heterogeneities  also influence the timing of the cancer progression  \cite{hallatschek1}. It is therefore useful to build explicitly spatial models to understand what to look for in clinical images and to better understand the spatial signatures of particular evolutionary dynamics.

  Although these spatial evolutionary aspects have only recently been explored in cancerous populations, many of the predictions of spatial models have been borne out in studies of microbial range expansions where  a  population of microbes grows into virgin territory (e.g., as a colony on a Petri dish). Here,   small number fluctuations and local fixation yield a sectoring phenomenon where initially mixed strains demix into uniform sectors containing a single strain over time \cite{DRNPNAS,hallatschekevo}.    The previously mentioned enhancement of deleterious mutations near population edges has also been verified via DNA sequencing of bacterial range expansions \cite{ExcoffierDeleterious1}.   Also, the  mutational meltdown we will consider in this paper has been demonstrated in yeast cell colonies, where a simple lattice model of the kind employed in this work successfully predicted the effects of the increased genetic drift in spatial populations \cite{Wahl}. Increasingly, results from such microbial populations and simple spatial evolutionary models are yielding insights into what may happen in cancerous populations \cite{FuscoJackpot,rangetumor}. The results presented here are also applicable to the microbial range expansions.

The evolutionary dynamics explored here (e.g., the motion and coarsening of the sectors of strains) has universal properties tying together a large class of systems including tumor growth, reaction-diffusion processes, granular material avalanches, and epidemic spreading \cite{Hinrichsen}.   For example, tumor shapes have been shown to have the same fractal boundary properties as films deposited by molecular beam epitaxy \cite{TumorMBE}. Therefore, many of the techniques originally developed to understand physical phenomena, such as the phase ordering of deposited binary films   \cite{drossel1,drossel2}, may be employed to understand the spatial evolutionary dynamics of microbial populations \cite{horowitzkardar}.   The universal properties of all of these systems include coarse-grained properties  such as the scaling of interface roughness with time \cite{KPZreview}, a quantity we will explore here for the interface between the bystander and the invader.     We may thus reasonably expect our results to hold generally, as we will be concerned with such coarse-grained properties.

 In cancerous tissue, current sequencing techniques have a limited ability to probe the spatial structure of the cancer cell population. Adaptation of sequencing techniques to spatially-distributed populations is important as spatial effects have been shown to significantly impact DNA sequencing data of cancerous cell populations \cite{SpatialCancerSequence}. Our study presents a complementary approach where we show that  physical cues such as the shape of an interface between competing cellular strains may indicate certain properties of the evolutionary dynamics of the tissue (e.g., the proximity to a mutational meltdown transition).  Such heuristic measures are useful in conjunction with DNA\ sequence information, which is often difficult to interpret and does not typically take into account the spatial structure of the cancerous population \cite{SpatialCancerSequence}.

 In this paper, we build a  model for how a mutating strain invades a non-mutating strain in both one- and two-dimensional habitats, which we call $d=1+1$ and $d=2+1$ evolutions, respectively.  The $+1$  indicates the time dependence.  Our focus here is on the competition between multiple strains within a population, so for simplicity we consider \textit{flat} habitats which do not change shape as the population evolves. For $d=1+1$, such a habitat may be a coast, a river bank, or a thin duct. For $d=2+1$, the strains may be in a microbial population growing on a flat surface or in an epithelial tissue. Another possibility is that the population in which the strains compete is the leading edge of a range expansion.   In this case, we assume  that the population growth is confined to a thin region at the edge, which remains flat during the range expansion. This approximation will hold as long as there is a sufficient surface tension keeping the population edge uniform which occurs in yeast cell colonies, for example \cite{yeastsurface}. However, if the population edge roughens over time, the roughening will generically change the genetic sector motion \cite{frey1}, an analysis of which is beyond the scope of the current work.   

Note that for a cellular population at the edge of a range expansion,  the $+1$ dimension represents the direction of the range expansion.  So, in other words, for $d=1+1$, the strains we study may live along a thin, effectively one-dimensional edge of a two-dimensional range expansion (e.g., a thin microbial colony grown on a Petri dish).  For $d=2+1$ evolution, the population  may  be the effectively two-dimensional, flat edge of a three-dimensional range expansion. A more realistic scenario is perhaps the $d=3+1$-dimensional case where a population embedded in three dimensions evolves in time with various strains within the population competing for the same space. Although we do not study this case specifically here, the lower dimensional cases provide intuitions for considering this scenario. Also, if the geometry of the three-dimensional population has a large aspect ratio, then our one and two (spatial) dimensional models will describe the behavior of cross sections of the population. A similar kind of dimensional reduction was recently employed for describing bacterial competition in three-dimensional channels \cite{channels}.

Previous work has shown that range expansions develop frontiers with enhanced roughness when the population is near a phase transition in its evolutionary dynamics (e.g., at a mutational meltdown transition  \cite{frey1}   or near the onset of mutualistic growth \cite{MOLMut}). In this work, we consider invasion frontiers which are markedly different as the invader grows into a surrounding population which may \textit{reinvade} if the invader growth rate decreases. So, the invasion front speed $v$ will depend on the relative growth rates of the two populations and may vanish or change signs.  In other words, the competition interfaces studied here have a variable speed, unlike a range expansion in which a population grows into a virgin territory with some particular growth rate. In this sense, the competition interface studied here is more similar to a \textit{range shift}, in which the population growth is limited by the environment \cite{ExcoffierDeleterious2}. We will see that in the case of $d=2+1$-dimensional expansions, the average speed $v$ of the invasion front will influence the interface roughness.

We will also show here  that, like the range expansion, the invasion frontier  develops an enhanced roughening at the mutational meltdown transition of the invader population. However, unlike a range expansion frontier, the roughening is more subtle, especially in the $d=2+1$-dimensional case where the relative growth difference  between the invader and bystander populations (and consequent invasion front speed $v$)  also influences the roughening dynamics. The invasion frontier does not maintain a compact shape, and isolated pieces of the invading population may pinch off and migrate into the surrounding ``bystander'' population, especially when the  growth rates of the two populations are nearly equal. In this paper we will discuss these issues and connect the shape of the undulating frontier to the evolutionary dynamics of the invading population.

The paper is organized as follows: In the next section, we present our lattice model for invasion by a mutating population for $d=1+1$ and $d=2+1$-dimensional cases.  In Section~\ref{sec:meltdown} we briefly review the nature of the mutational meltdown transition that may occur in  the unstable invading population. In Section~\ref{sec:phases}   we study the survival probability of the invading strain and construct phase diagrams characterizing whether or not the invasion is successful as a function of the internal mutation rate $\mu$ and the selective advantage of the various cellular strains. In Section~\ref{sec:roughening}, we analyze the roughening invasion front near the mutational meltdown transition for the invading population. Finally, we  conclude with a discussion of the implications of our results in Section~\ref{sec:conclusion}.

\section{Model}

We  consider a simple lattice model,  in the spirit of the Domany-Kinzel cellular automaton   \cite{RadDK,DK}, of invasion of a stable population by a mutating invader consisting of two species, a fast-growing and a slow-growing strain into which the fast-growing one can mutate. We  set the fast-growing strain growth rate to unity $\Gamma_f=1$ without loss of generality, so that time is measured in generation times $\tau_g$ of the fast-growing strain. The slow-growing strain within the invader population will have growth rate   $\Gamma_s = 1-s$, where  $0 \leq s<1$ is a measure of the deleterious effect of the mutation.     In a   tumor or microbial colony, we know that the initial cluster of  growing cells may encounter other cells (e.g., surrounding healthy tissue or competing microbial strains). 
So, we have a third species representing the ``bystander'' population. The bystander will not mutate, but will be able to displace the mutating population via cell division. We set the bystander growth rate to $\Gamma_b=1+b-s$, with $b$ the selective advantage of the bystander over the slow-growing invader strain.

The internal dynamics of the invading population (i.e.,  the mutation rate $\mu$ and selection parameter $s$) will influence how the invader interacts with the bystander strain, with increasing $\mu$  or $s$ leading to an overall fitness decrease for the invading population, as might happen in a cancerous tissue during a course of therapy that increases the deleterious mutation rate of the cancerous cells.   We focus on the region between the mutating population and the bystander, which we call the invasion front. As we will show, when the invader is close to mutational meltdown, the invasion front develops an enhanced ``roughness.''

\begin{figure}[htp]
\centering
\includegraphics[width=0.6\textwidth]{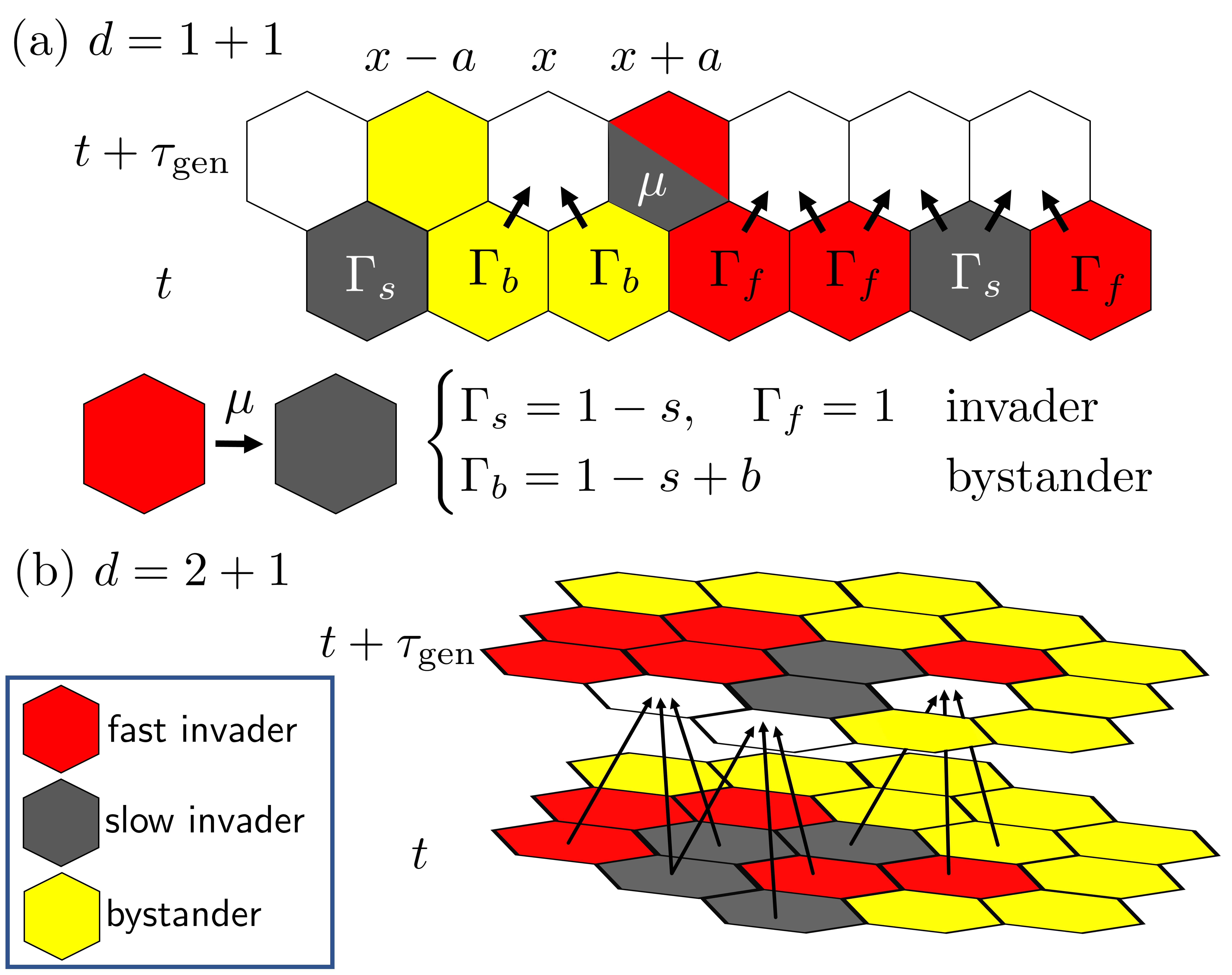}
\caption{\label{fig:bystander}(a) Update rules for the bystander model for a   population in $ d=1+1$ dimensions. Each generation is evolved by allowing for two cells from the previous generation to compete for an empty lattice site, as shown by the arrows. The probability of occupation by a cell of a type $i=s,f,b$ is proportional to its growth rate $\Gamma_i$, where $s$ is the slow growing black strain, $f$ is the fast growing red strain, and $b$ is the yellow bystander. If a red (fast-growing) cell is placed in the empty spot, then it in addition has a probability $\mu$ of mutating to the slower-growing black strain. (b) For a two-dimensional population $(d=2+1)$ the generations are evolved on staggered triangular lattices, as shown. This time, \textit{three} cells compete to divide into empty lattice sites. Otherwise, the dynamics are the same as the $d=1+1$ case.     }
\end{figure}

The specific lattice model rules are as follows: In both one- and two-dimensional population scenarios, we consider a three-strain model in which a single ``bystander'' strain (yellow cells in Fig.~\ref{fig:bystander}) grows in the presence of a fast-growing invading strain (red cells in Fig.~\ref{fig:bystander}) that can mutate to a more slowly growing strain (black cells in Fig.~\ref{fig:bystander}). These cells occupy a single lattice location, as shown in Fig.~\ref{fig:bystander}. During each generation (cell division) time $\tau_g$, the lattice of cells is regenerated by allowing for adjacent cells to compete and divide into empty sites representing the next generation of cells. In a range expansion context, these empty sites would represent unoccupied territory at the frontier. Alternatively, these updates can represent a turnover of cells due to birth and death within the population.  After all empty sites in the next lattice have been filled (rows for $d=1+1$ and sheets for $d=2+1$ as shown in Fig.~\ref{fig:bystander}(a) and (b), respectively), the process can be repeated, generating a sequence of successive generations of the population (or, alternatively, a moving frontier of a range expansion). Note that each time a red, fast-growing cell is chosen to occupy an empty site, it mutates to a slower-growing black strain with probability $\mu$.

\begin{figure}[htp]
\centering
\includegraphics[width=0.5\textwidth]{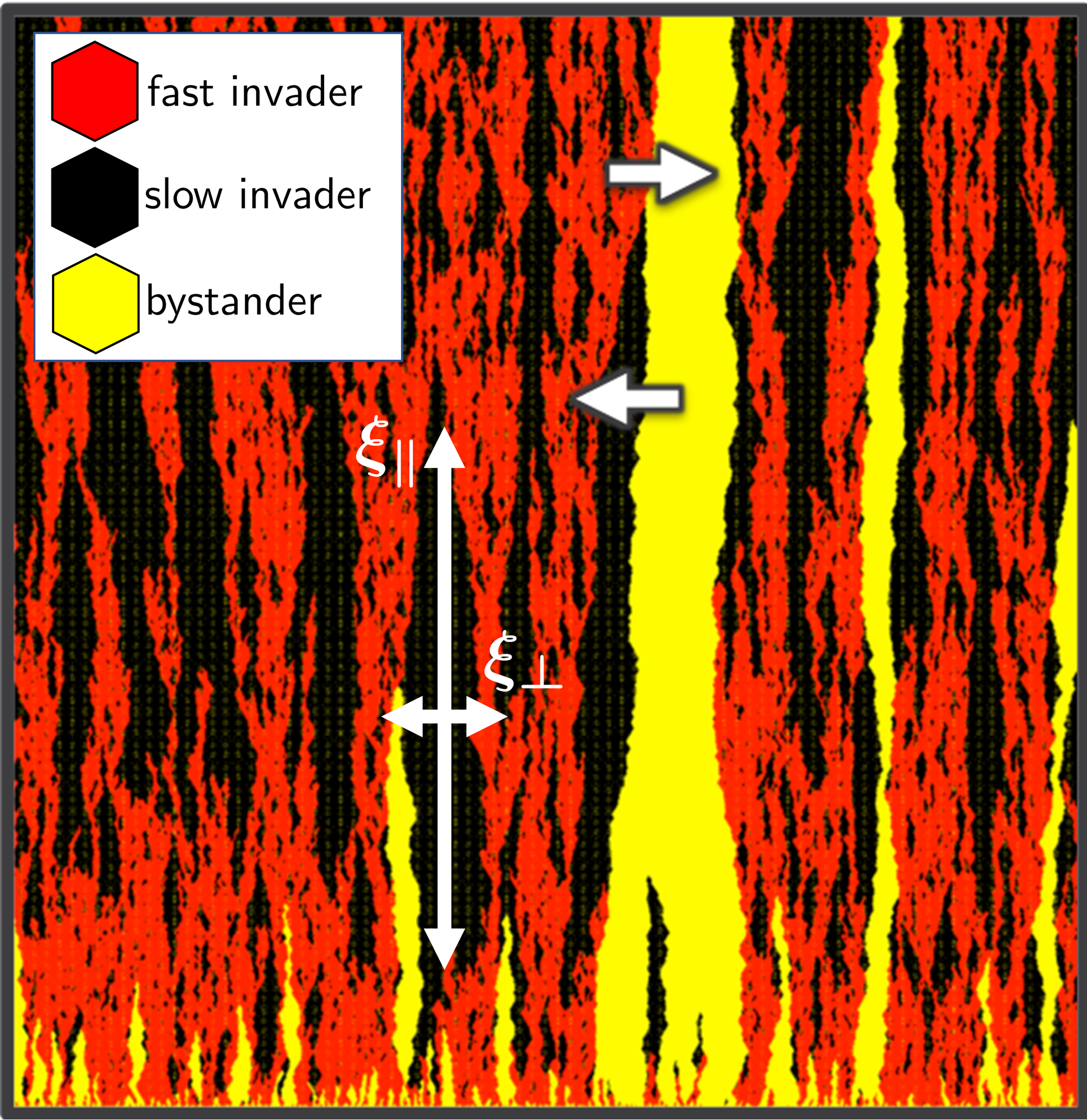}
\caption{\label{fig:alternatingRW} A $d=1+1$ simulation of a red/black mutating population invading a bystander yellow one. Here, the yellow strain grows faster than the black strain but slower than the fit red strain. The invasion front between the black/red population and the bystander strain can be characterized by a random walk with alternating bias. The yellow strain invades the black patches and is invaded by the red patches. The sizes of the red and black patches are controlled by the internal dynamics of the black/red strain. We illustrate the characteristic sizes $\xi_{\perp}$ and lifetime $\xi_{\parallel}$ of the black patches.} 
\end{figure}

 So, our model contains just three parameters: the deleterious effect of the mutation $s$, the selective advantage $b$ of the bystander population over the slow-growing strain, and the mutation rate $\mu$.  We will be interested in the regime $0<\Gamma_s \leq \Gamma_b \leq \Gamma_f$, for which $0 \leq s < 1$ and $0 \leq b \leq s$. In this case, the bystander either invades the slow-growing strain or is invaded by the fast-growing strain, as illustrated in Fig.~\ref{fig:alternatingRW} for a $d=1+1$ simulation. Note that this reinvasion by the bystander population makes the invasion frontier markedly different from, say, a range expansion frontier.  In a range expansion frontier, the range expansion always moves in one direction according to the growth rate of the total population. Here, the interface between the bystander and invader can move in different directions or even remain, on average, stationary. We will see that this aspect will be important when studying the roughness of the interface.

Our parameterization allows us to tune the dynamics of the black/red mutating invader  population separately. As we will analyze in the next section, the invader has an internal ``mutational meltdown'' at which the fast-growing red strain is removed from the population due to mutation. This occurs for $\mu \gtrsim s^2$ in $d=1+1$ dynamics and $\mu \gtrsim s \ln s$ in $d=2+1$ dynamics ($\mu > s$ in well-mixed populations).  Note that an important limitation of our model is that we assume cells divide into adjacent spots on the lattice so that cell motility is essentially absent (apart from the short-range cell rearrangements occurring due to the cell division). This is a reasonable approximation for certain microbial populations such as yeast cell colonies \cite{Wahl} or small, avascular solid tumors where cells primarily proliferate \cite{migration}.

\section{Mutational meltdown \label{sec:meltdown}}

\begin{figure}[htp]
\centering
\includegraphics[width=0.8\textwidth]{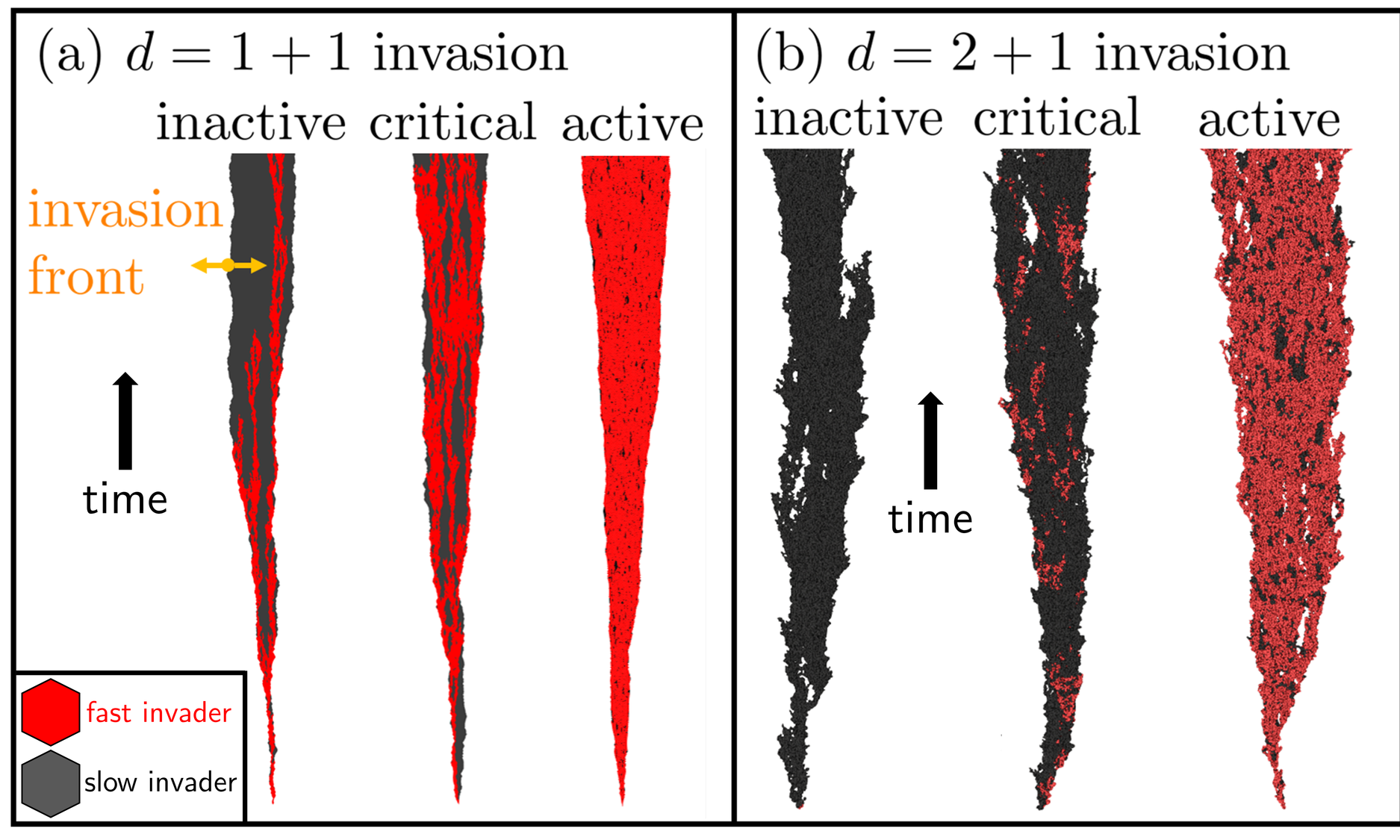}
\caption{\label{fig:sectors}Simulated sectors of a black/red mutating population invading a bystander population (surrounding white area), initialized by a single red cell in a (a) one-dimensional and (b) two-dimensional population. The populations are evolved for about 100 generations, with the time direction indicated. In (a), we indicate the motion of the invasion front (which in this case is a point) between the two populations.  In (b), the invasion front would be the complicated boundary between the black/red population and the white space at each time slice $t$ along the vertical direction. The phases of the internal dynamics of the invading population (inactive, critical, and active phases) are indicated. In the inactive phase, the red, fast-growing mutant is lost from the population over time. As the invader population transitions from the inactive to the active phase in which the red strain is maintained, the invasion front exhibits enhanced undulations.}
\end{figure}

 Let us first focus on the invading population and perform a simple analysis of the internal dynamics. The invader consists of two strains: one fast-growing red strain and a second slow-growing black strain   into which the fast-growing strain mutates with  rate $\mu$ per cell per generation.  In the parameter space  $(\mu,s)$, we find two distinct phases \cite{RadDK}: In one phase, the fraction of the  fast-growing strain remains positive after many generations, $\rho_{f}(t\rightarrow\infty) > 0$; we call this phase the \emph{active phase}. In the other phase, called the \emph{absorbing} or \emph{inactive phase}, the fast-growing strain eventually completely dies out and $\rho_{f}(t\rightarrow\infty) = 0$. There is a line of continuous phase transitions $(\mu^*, s^*)$ which defines the boundary between the two phases. Examples of these phases, and the critical region ($\mu \approx \mu^*$, $s \approx s^*$), are shown in Fig.~\ref{fig:sectors} where we have removed the bystander population in order to see the internal dynamics of the invader. 

 We can understand this transition in a well-mixed population (a mean-field approach). Consider just the invading, mutating population. The fraction $\rho_f \equiv \rho_f(t)$ of the fast-growing strain within the mutating population  changes according to:
\begin{equation}
\frac{d \rho_f}{dt}=s \rho_f(1-\rho_f)-\mu \rho_f,
\end{equation}
which approaches $\rho_f(t\rightarrow \infty) = 1-\mu/s$ for $\mu<s$, and $\rho_f(t \rightarrow \infty)= 0$ for $\mu>s$. The line $\mu =s$ is our set of critical points $(\mu^*,s^*=\mu^*)$. For a spatially distributed population, small number fluctuations or ``genetic drift'' dramatically alters the shape of the phase boundary:  The phase transition occurs for $\mu^* \sim (s^*)^2$ in one-dimensional populations (such as at the edge of a growing biofilm \cite{Wahl}) and $\mu^* \sim s^* \ln (s^*)$ for two-dimensional populations  \cite{lavrent1}. This phase transition, called a ``mutational meltdown,'' is known to belong to the directed percolation (DP) universality class \cite{Hinrichsen}. For the particular lattice model we consider here, this has been explicitly verified   \cite{RadDK} by mapping the model to the well-studied Domany-Kinzel cellular automaton \cite{DK}. The efficacy of this simple model has been  demonstrated in a synthetic yeast strain, for which the parameters  $\mu$ and $s$ could be tuned over a broad range encompassing the DP phase transition \cite{Wahl}.

Note that when the population approaches the mutational meltdown transition, the slow-growing black strain within the population begins to take over. In the active phase in Fig.~\ref{fig:sectors}, the black strain makes finite-sized, small patches within the red population. Then, as we approach the transition, the black strain patch sizes diverge. In the critical regime, the average patch size becomes infinite. Then, in the ``inactive'' phase, the red strain will eventually die out completely, leaving behind just the slowly growing black strain.    We shall see that it is this divergence of the black strain patch size near the transition which is responsible for enhanced invasion front roughening.  

The two-species model may be generalized to include an arbitrary number of possible mutations, and such models have been shown to exhibit critical behavior that deviates from the DP universality class, but the loss of the fittest mutant in the population is still well-described by DP \cite{lavrent1}.   The multi-species generalization has many additional interesting phenomena such as multi-critical behavior \cite{UCDP}, which would allow for interesting extensions of the work presented here.  In this paper, for simplicity, we shall focus on the fittest mutant in an invading population with just two species. The fittest strain could represent, for example, a driver mutation which has swept through a cancerous tissue. The driver strain could then acquire deleterious mutations over time with rate $\mu$.  We will focus here on just the initial loss of fitness, characterized by a single mutation to the slower growth rate $\Gamma_s=1-s$.

\section{Invasion probabilities \label{sec:phases}}

We now construct a phase diagram for successful invasion of the bystander strain by the mutating invader (Fig.~\ref{fig:threespeciesPD}). We initialize a well-mixed population of equal parts of the mutating red and  bystander yellow strains $(\rho_b=\rho_f=1/2)$ on the lattice, and then calculate the density $\rho_m=\rho_f+\rho_s$ of the mutating red/black population at long times $t$. If $\rho_m \rightarrow 0$ and the red/black population dies out at long times, then the ``invasion'' is unsuccessful. Otherwise, $\rho_m$ approaches a non-zero value and the invasion succeeds.  The results are shown in  Fig.~\ref{fig:threespeciesPD} for  $d=1+1$ and $d=2+1$, where we see the two distinct phases. We also include the phase boundary $\mu=s-b$  for the well-mixed population (blacked dashed line in Fig.~\ref{fig:threespeciesPD}), which we derive in the next subsection. Note how far away the well-mixed population transition line is from the actual transition in a spatial population. The genetic drift associated with the spatial populations suppresses the invasion by the red/black mutating population.

We also know that as we approach this mutational meltdown transition  at $\mu=\mu^*$ for a fixed $s$ [vertical dashed lines in Fig.~\ref{fig:threespeciesPD}], the characteristic sizes $\xi_{\perp}$ and the characteristic lifetimes $\xi_{\parallel}$ of black, slow-growing strain clusters diverge according to $\xi_{\perp} \sim \Delta^{-\nu_{\perp}}$ and $\xi_{\parallel} \sim \Delta ^{-\nu_{\parallel}}$, where $0<\Delta<1$ is the distance from the phase transition in the $(\mu,s)$ plane and  $\nu_\perp $ and $\nu_\parallel $ are critical exponents associated with directed percolation ($\nu_{\perp}\approx  1.097$, $\nu_{\parallel} \approx  1.734$ for $d=1+1$ and $\nu_{\perp} \approx 0.734$, $\nu_{\parallel} \approx 1.295$ for $d=2+1$ \cite{Hinrichsen}). We illustrate the sizes $\xi_{\perp,\parallel}$ in Fig.~\ref{fig:alternatingRW}.  The black patches will interact differently with the bystander than with the red patches of the fast-growing strain. As the patch sizes $\xi_{\parallel,\perp}$ diverge (when $\Delta \rightarrow 0$), they would have a more pronounced effect on the invasion dynamics. In particular, there will be larger regions over which either the yellow strain invades a black patch, or a red patch invades the yellow bystander. This will increase the amount of ``wiggliness" of the invasion front between the bystander and the mutating red/black population. We will see in the following that there is  a significant enhancement of the roughness as we approach the mutational meltdown transition. 

\subsection{Mean field analysis}

To understand the behavior of this three-species model, we first briefly describe what happens in a well-mixed (mean-field) context. Consider the  time-evolution of the fractions $\rho_f,\rho_s,\rho_b$ of the fast-growing, slow-growing, and bystander strains, respectively. For a fixed total population size, we have  $\rho_f+\rho_s+\rho_b=1$.  Given our growth rates  $\Gamma_f=1$,   $\Gamma_s = 1-s$, and $\Gamma_b = 1-s+b$, we may define corresponding selection coefficients characterizing the competition between pairs of strains:
\begin{eqnarray}\label{eq:selcoeff}
s_{bs} & = & \frac{\Gamma_b - \Gamma_s}{(\Gamma_b + \Gamma_s)/2} = \frac{2b}{2-2s+b}\nonumber\\
s_{fb} & =& \frac{\Gamma_f - \Gamma_b}{(\Gamma_f + \Gamma_b)/2} = \frac{2(s-b)}{2-s+b}\\
s_{fs}&  = & \frac{\Gamma_f - \Gamma_s}{(\Gamma_f + \Gamma_s)/2} = \frac{2s}{2-s},\nonumber
\end{eqnarray}
with selection parameters $0<s<1$ and $0<b<s$. In terms of these selection coefficients, the  equations for the time-evolution of the bystander and fast-growing strain fractions $\rho_{b,f} \equiv \rho_{b,f}(t)$  in a well-mixed population are
\begin{equation}\label{eq:meanfield}
\left\{
\begin{array}{@{\kern2.5pt}ll}
\partial_t\rho_b = s_{bs}\rho_b\left(1-\rho_b-\rho_f\right) - s_{fb}\rho_f\rho_b\\
\partial_t\rho_f = s_{fs}\rho_f\left(1-\rho_b-\rho_f\right) + s_{fb}\rho_f\rho_b - \mu\rho_f
\end{array} \right. .
\end{equation}
If $\rho_b = 0$, we recover the two-species dynamics of the invader population with a directed percolation-like process between the fast-growing and slow-growing strains. We can also verify that there is no sensible stable fixed point where both the bystander population and the invader coexist. Instead, if   $\mu> s-b$, then the bystander will sweep the total population and $\rho_b(t) \rightarrow 1$ with increasing time $t$. Otherwise, if $\mu < s-b$,  the invasion by the mutating population is successful  and we find $\rho_b(t) \rightarrow 0$ over time. Moreover, if $\mu>s$, we get a collapse of the fast-growing strain [$\rho_f(t) \rightarrow 0$], and then the  bystander strain will win out since $\Gamma_s < \Gamma_b$. So, the mutational meltdown transition of the invader population occurs when $\mu=s$ in this mean field analysis.

The mean-field analysis tells us that we should expect a critical \textit{surface} in the $(\mu,s,b)$ parameter space given by $\mu=s-b$ separating a region of successful   $(\mu< s-b)$ or failed $(\mu> s-b)$ invasion of the bystander strain by the mutating invader (which itself may undergo a mutational meltdown when $\mu>s$). As we have already seen, the spatially-distributed populations also have this critical surface but the enhanced genetic drift  \textit{suppresses} the phase space for successful invasion. To add the effects of genetic drift and the spatial distribution of the population to Eq.~\eref{eq:meanfield}, we would have to incorporate a spatial diffusion term $\nabla^2 \rho_{b,f}$ in each of the equations and   stochastic noise terms describing the birth/death dynamics (see \cite{lavrent1} for a more detailed description).
These additional terms significantly modify our mean field equations and introduce different phenomena, such as propagating waves (moving population interfaces) which we will analyze in Section~\ref{sec:roughening}.

\begin{figure}[htp]
\centering
\includegraphics[width=0.55\textwidth]{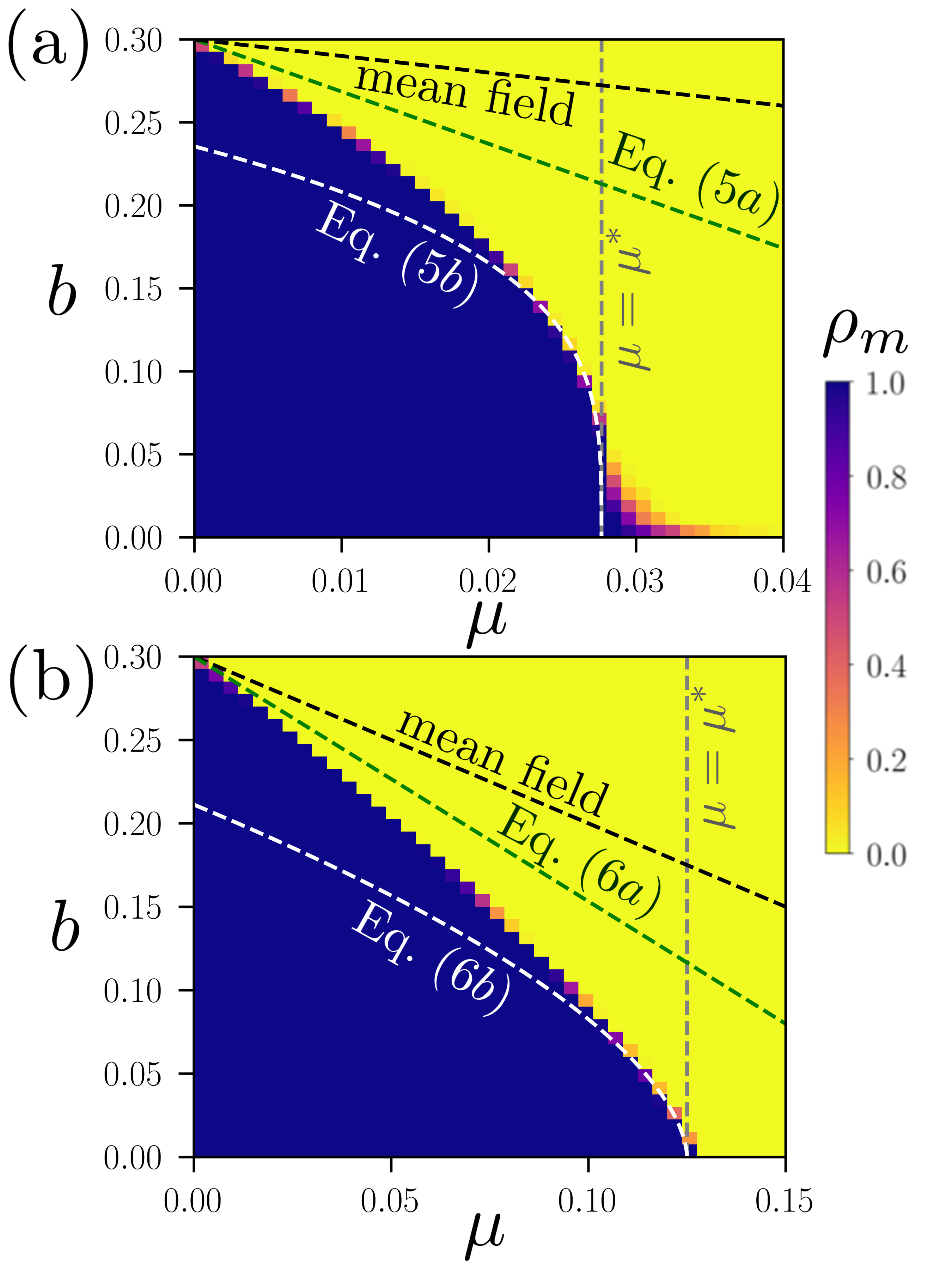}
\caption{Phase diagrams for  (a) the $d=1+1$  case and (b)  the $d=2+1$ case, calculated by initializing a well-mixed population of the red and yellow strains and evolving the whole population for $t \approx 10^6$ generations. In (a) we use a one-dimensional population of  $L=5000$ cells and average over 256 runs of the evolution. In (b) we have a two-dimensional population with $L^2$ cells, where $L=500$. Here we average over 40 evolution runs.  In both cases we set $s=0.3$.  After evolving for $10^6$ generations, we calculate the red/black mutating fraction of the total population: $\rho_m=\rho_f+\rho_s$.  In each phase diagram, the black dashed line corresponds to the mean field prediction $\mu=s-b$. The green and white dashed lines correspond to the improved predictions [see Eqs.~\eref{eq:improvedmf1}, \eref{eq:improvedmf2}, \eref{eq:improvedmf3d1}, \eref{eq:improvedmf3d2}] for $\mu \ll \mu^*$ and $\mu \approx \mu^*$, respectively, that take into account the spatial structure of the population. We also indicate the line $\mu=\mu^*$ along which we find a mutational meltdown transition within the invading red/black population.  \label{fig:threespeciesPD}}
\end{figure}

We can get a better approximation to the critical line for the $d=1+1$ case than that given by the mean field theory by considering a single domain wall. We expect the total length along the domain wall to be split into sections of average length $\ell_s$ where the slow-growing strain competes with the bystander  and sections of average length $\ell_f$ where the fast-growing strain competes with the bystander. During each generation time $\tau_g$ the domain wall can move by a single cell length $a$. So, in the fast-growing strain segments, we expect the fast-growing clusters to out-compete the bystander and protrude by an amount $ \tau_g\ell_f (s-b)/a$, with $\tau_g$ the generation time. Similarly, we expect the slow-growing clusters to be out-competed by the bystander and recede by an amount $ \tau_g\ell_s b/a$. At the phase transition, we expect these competitions to cancel each other out as the mutating invading population is \textit{barely} able to invade the bystander in this case. Hence,  we should have $ \ell_f (s-b) \approx \ell_s b $ so that
\begin{equation} b  \approx  \frac{\ell_fs}{\ell_s+\ell_f} \approx  \phi_f \, s,    \label{eq:deltas}
\end{equation}
where $\phi_f$ is the red-cell (fast-growing) fraction of the mutating invader population.   We now can use Eq.~\eref{eq:deltas} to predict the critical line in $(\mu,b)$-space  for a fixed $s$ in two limiting cases: $\mu \ll \mu^*$ and $\mu \sim \mu^*$, where $\mu^*$ is the critical value for $\mu$ for the specific fixed value of $s$ at which we get the mutational meltdown transition within the red/black invading population.

To complete the derivation, we just need an estimate for the fraction of fast-growing strain $\phi_f$. First, when $\mu \ll \mu^*$, the invader population is in the ``active'' phase, and the patches of black slow-growing strain are small and rarely collide, as shown in Fig.~\ref{fig:sectors}. In $d=1+1$-dimensions, the boundaries of these black patches are well-described by pairs of random walkers, yielding an estimate   $\phi_f \approx 1-A_1\mu/s^2$ \cite{RadDK,otwinowski,hallatschekevo}. The amplitude $A_1$ is model-dependent, and we have $A_1 \approx 0.5$ for this Domany-Kinzel-type model, consistent with previous results \cite{lavrent1}.  As $\mu \rightarrow \mu^*$, however, the fast-growing strain vanishes  $(\phi_f \rightarrow 0)$, and we have to make another approximation. From the random walk model, we expect that $\phi_f$ vanishes when  $\mu=\mu^* \sim s^2$. Then, when $\mu \approx \mu^*$, we would be near a directed percolation (DP) phase transition where  $\phi_f$ serves as an order parameter. The order parameter vanishes according to  $\phi_f \approx A_2(\mu^*-\mu)^\beta$, where $A_2$ is an amplitude that will depend on $s$ and $\beta\approx 0.276$ is a DP critical exponent \cite{Hinrichsen}.  We may now use Eq.~\eref{eq:deltas} to make an estimate of the critical value of $b$  for $d=1+1$:
\numparts\begin{eqnarray} 
 b &=   
      s(1-A_1\mu/s^2) \qquad   & ( \mu \ll \mu^*) \label{eq:improvedmf1} \\
   b& =     sA_2(\mu^*-\mu)^\beta & ( \mu \approx \mu^*)  \label{eq:improvedmf2}
\end{eqnarray}\endnumparts
 where $A_1$ and $A_2$ can be calculated numerically from separate simulations of the two-species model. These improved estimates  are plotted onto the phase diagrams in Fig. \ref{fig:threespeciesPD}(a) (green dashed line for the $\mu \ll \mu^*$ case and white dashed line for the $\mu \approx \mu^*$ case).

For $d=2+1$-dimensional evolutions, the situation is more complicated because the patches of the invader strain no longer have a compact shape describable by a simple random walk [see Fig.~\ref{fig:sectors}(b)]. However, we   expect that the bystander population may reinvade the invading mutating population when $b> s \phi_f$ because, much like in the $d=1+1$ case, the average growth rate of the invader strain is approximately $\overline{\Gamma} \approx \phi_f \Gamma_f+(1-\phi_f) \Gamma_s=1-s+\phi_f s$. The bystander strain has growth rate $\Gamma_b = 1-s+b$, so we see that the growth rates are equal when $b = \phi_f s$. We now just need estimates for $\phi_f$ for $d=2+1$.  When $\mu \ll \mu^*$, previous work \cite{lavrent1} has shown that $\phi_f \approx1-A_3 \mu \ln (s/s_0)/s $, with $A_3 \approx 0.3 $ and $s_0 \approx40$ some model-dependent parameters. Conversely, when\ $\mu \approx \mu^*$, we again find a DP transition with $\phi_f \approx A_4(\mu^*-\mu)^{\beta}$ with critical exponent $\beta \approx0.584$ for $d=2+1$ \cite{Hinrichsen}.
The corresponding estimates are for $d=2+1$:
\numparts\begin{eqnarray} 
 b &=   
      s[1-A_3 \mu \ln (s/s_0)/s] \qquad   & ( \mu \ll \mu^*) \label{eq:improvedmf3d1}\\
   b& =     sA_4(\mu^*-\mu)^\beta & ( \mu \approx \mu^*)  \label{eq:improvedmf3d2}
\end{eqnarray}\endnumparts
These two approximations are plotted in Fig.~\ref{fig:threespeciesPD}(b), with $\mu \ll \mu^*$ in the green dashed line and $\mu \approx \mu^*$ in the white dashed line.

The phase diagrams in Fig. \ref{fig:threespeciesPD} were constructed with simulations using mixed initial conditions; the first generation of cells on the lattice were populated by an even mixture of fast-growing (red) cells and bystander (yellow) cells. These phase diagrams are heat maps corresponding to the density $\rho_m$ of the mutating population (red/black strains) after many generations. Our simulations were performed for $t \approx 10^6$ generations, which yields the steady state solution for the mutating population fraction $\rho_m$ for the vast majority of points on the phase diagram in Fig.~\ref{fig:threespeciesPD}, except for points very near the phase transition line. Note that our improved mean field estimates based on directed percolation and the random walk theory (white and green dashed lines, respectively) do a reasonable job of approximating the shape of the phase boundary, especially when $\mu \approx 0$ and our system reduces to a simple competition between fast-growing red cells and bystander yellow cells. The directed percolation approximation works better near $\mu \approx \mu^*$, where we find the mutational meltdown transition of the invader population which is in the directed percolation universality class.

\begin{figure}[htp]
\centering
\includegraphics[width=0.8\textwidth]{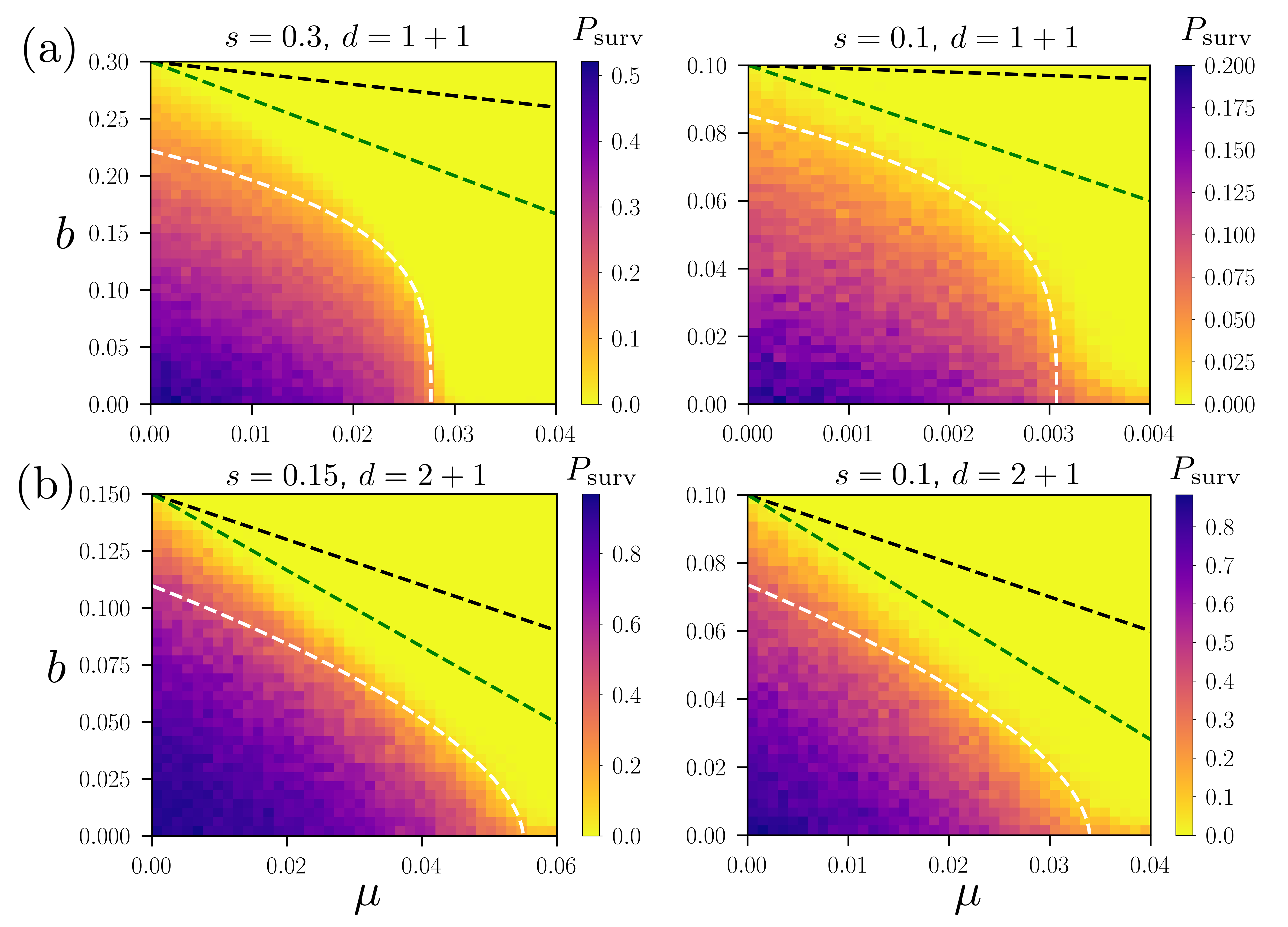}
\caption{Long-time survival probabilities $P_{\mathrm{surv}}$ of clusters generated from a single mutating red cell in a yellow bystander population for  (a) $d=1+1$-dimensional range expansions for $s=0.3$ (left panel) and $s=0.1$ (right panel), after $t=10^6$ generations on a lattice with size $L=5000$ cells averaged over 128 runs; and (b) $d=2+1$-dimensional range expansions for $s=0.15$ (left panel) and $s=0.1$ (right panel), after $t=2\times 10^3$ generations on a lattice with size $L^2$ cells with $L=500$, averaged over 256 runs. We show the expected transition shape near the $\mu\approx \mu^*$ DP transition in the white dashed line [Eqs.~\eref{eq:improvedmf2},\eref{eq:improvedmf3d2}]. The black dashed line is the transition position for a well-mixed population. The green dashed line is an improved mean-field estimate of the transition discussed in the main text [Eqs.~\eref{eq:improvedmf1},\eref{eq:improvedmf3d1}].  } \label{fig:PDcompare}
\end{figure}

 Another biologically interesting quantity to look at is the survival probability $P_{\mathrm{surv}}$ of the progeny of a single red cell invader in a population of yellow bystander cells as $t \rightarrow \infty$. Such a survival probability would represent the probability of tumor invasion, for example, from a single mutated cell (i.e, a cell with a newly-acquired driver mutation) within an otherwise healthy population.  If the bystander is replaced by the slow-growing strain and we have a two-species evolution, then the evolution will be exactly the same as a directed percolation with a ``single-seed'' initial condition. We would then have $P_{\mathrm{surv}} \propto \rho_f$ due to the rapidity reversal symmetry of directed percolation \cite{Hinrichsen}. In other words, the survival probability  tracks the behavior of the fraction $\rho_f$ of the fast-growing strain in a \textit{different} simulation where the initial condition is a \textit{well-mixed} population (or a population of just the mutating, fast-growing strain). In the three-species model we consider, there is no rapidity symmetry due to the presence of the bystander strain. Nevertheless, we expect that the survival probability $P_{\mathrm{surv}}$ vanishes on the same critical surface as the fraction $\rho_m$ (plotted in Fig.~\ref{fig:threespeciesPD}) because the invader strain will not be able to invade if the fast-growing strain is lost from the population.     We show in  Fig.~\ref{fig:PDcompare} the survival probability $P_{\mathrm{surv}}$ at long times, which does indeed vanish at approximately the same place as $\rho_m$ in Fig.~\ref{fig:threespeciesPD}. So, the approximations we used to estimate where $\rho_m$ vanishes serve as good predictors of the transition of the survival probability, as well. We also show the phase boundary at a smaller values of $s$ ($s=0.1$) in the right panels of Fig.~\ref{fig:PDcompare}. Note that our   estimates work for the phase boundary in this case, also.

\section{Roughening invasion fronts \label{sec:roughening}}

We now   study the shape of the interface between the mutating and bystander populations. When either of the populations is invading the other, the invasion front behaves as a noisy Fisher-Kolmogorov-Petrovsky-Piscounov wave \cite{fisher,KPPeq}.  Most previous studies of such waves have focused on competition between two homogeneous populations or the range expansion of a population into virgin territory. The noise plays a crucial role here \cite{korolevfisher}, strongly modifying, for example, the wave speed. Also, in the (exactly soluble \cite{doeringexact}) $d=1+1$ case, there is a diffusive wandering of the front around its average position.

For $d=2+1$, the situation is more complicated, but generally the noisy wave front will have a characteristic roughening. This roughening   falls in the Kardar-Parisi-Zhang (KPZ) universality class \cite{KPZ}, although observing the predicted scaling behavior of this class is challenging for  noisy Fisher waves \cite{MoroFisher,SaarloosFisher}. For example, for the KPZ class of interfaces, the characteristic size $\sigma_w$ of the interface should grow as $\sigma_w \propto t^{1/3}$. However, a basic analysis of the noisy Fisher waves \cite{benavraham} is more consistent with $\sigma_w \propto t^{0.272}$, which is also what we observe in our model. Although this apparent discrepancy has been explained, the proper recovery of the KPZ exponents requires a deeper analysis outside the scope of the current paper  \cite{MoroFisher}. So, for our simulations, we will find consistency with previous analyses of noisy Fisher waves and leave the more extensive analysis of the interface shape scaling for future work. Also, as the speed of the invasion goes to zero, we expect a transition to a different, ``voter-model'' \cite{clusteringvoter} interface coarsening behavior as both the mutating and the bystander populations become stable and do not invade each other (on average). The interface roughens in a different way in this ``critical'' case, with the characteristic size $\sigma_w$ of the interface increasing diffusively as $\sigma_w \propto t^{1/2}$. We will observe such crossovers in our simulation results. 

We discuss these issues in more detail below and show that our model exhibits a range of behaviors depending on the invasion speed and the proximity of the mutating population to the meltdown transition. These invasion waves are examples of ``pulled'' wavefronts \cite{OGPushedPulled}, which are driven by the growth (invasion) at the leading edge of the wave. Various aspects of such wave fronts are reviewed in, e.g., Ref. \cite{SaarloosReview}. We shall see  in the following that adding mutations to one of the populations significantly modifies the expected pulled front wave behavior and, in the $d=1+1$ case, introduces a super-diffusive wandering of the interface.
The $d=2+1$ case presents an even richer set of behaviors depending on the mutation rate and relative fitness of the mutating and bystander populations.
Our purpose here will not be the particular value of scaling exponents, but rather general features of the roughening dynamics such as a change in roughening due to the internal evolutionary dynamics of the invading strain.
\subsection{$d=1+1$-dimensional invasions}

\begin{figure}[htp]
\centering
\includegraphics[width=0.5\textwidth]{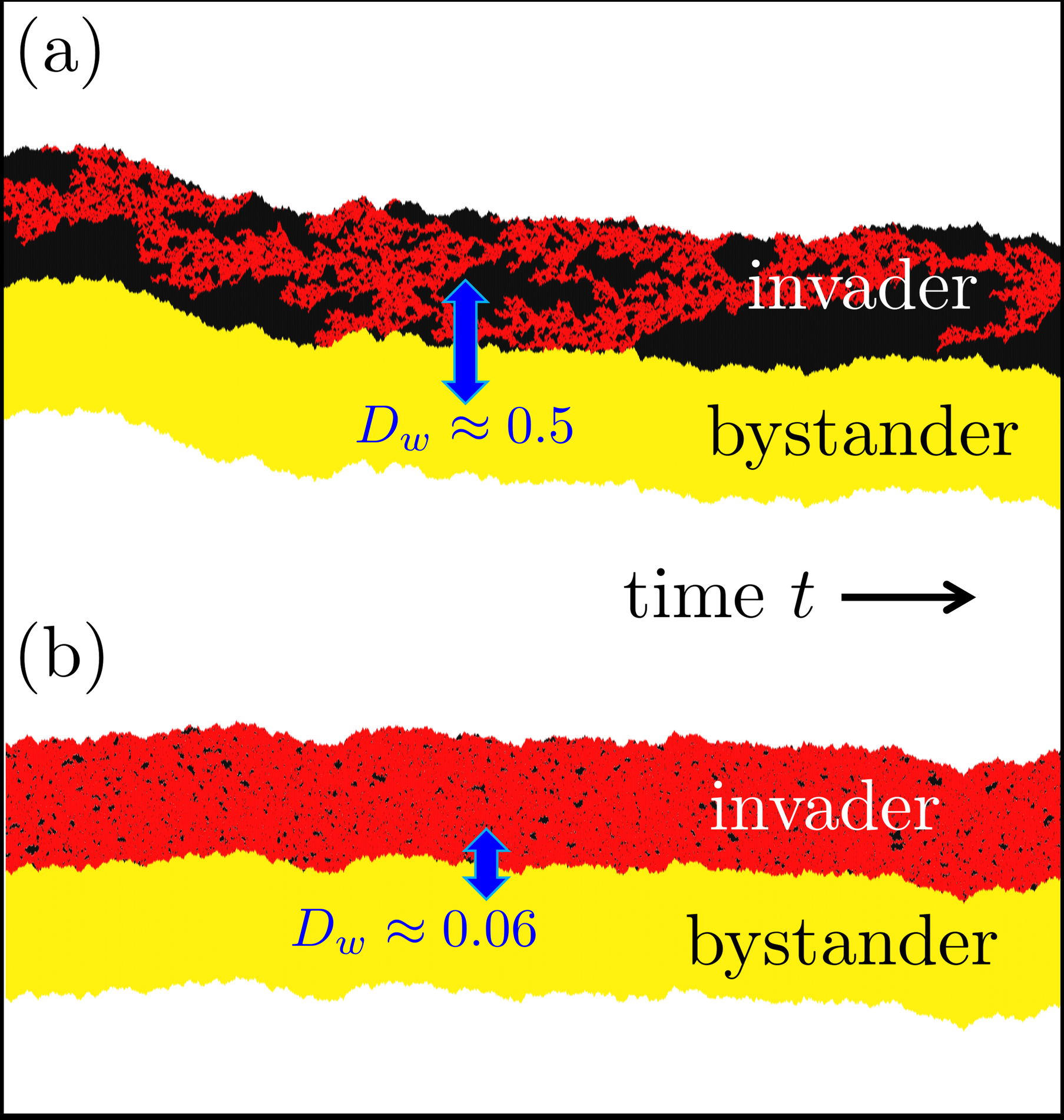}
\caption{\label{fig:dw}A picture of the domain wall in a $d=1+1$-dimensional evolution between the invading black/red population and the bystander yellow population (a) near the  two-species (DP) phase transition ($\mu=0.02765$, $b=0.045$, $s=0.3$) and  (b) far from the mutational meltdown DP phase transition ($\mu=0.005$, $b=0.2775$, $s=0.3$). We see that the roughness of the domain wall becomes enhanced for (a) with a domain wall diffusion coefficient $D_w \approx0.5 $ as compared to $D_w \approx0.06 $ for (b) (calculated using the time-averaged squared displacements of the wandering interface in the corresponding figures \cite{diffusivities}). The width of the population is $L=300$ cells, and we show the evolution over $10^4$ generations. This long evolution time allows for an observation of the domain wall wandering. However, as a result, the cells are compressed along the time direction in this figure.}
\end{figure}

In $d=1+1$, domain walls between the bystander  and the invading populations can be characterized by a random walk with alternating bias (when $\Gamma_s<\Gamma_b < \Gamma_f$) as the bystander will invade the slow-growing species and be invaded by the fast-growing species within the mutating population. As we approach the mutational meltdown transition, the average size of clusters of the slow-growing strain will diverge as expected from the directed percolation transition.  In Fig. \ref{fig:dw} we see a comparison of two domain walls for $d=1+1$. At the bottom of the figure, Fig.~\ref{fig:dw}(b), we see a domain wall where the mutating red/black population is far from the two-species phase transition. In this case, the black patches in the population are small and do not influence the motion of the invasion front much. Conversely, in Fig.~\ref{fig:dw}(a), we see a  domain wall with the mutating population near a mutational meltdown. In this case, there is an enhancement of the ``roughness'' of the domain wall as the large black patches create more regions of alternating bias in the domain wall between the yellow bystander and the red/black invading population.

To obtain a more quantitative estimate of this roughening effect, we set up a simulation with initial conditions that include a sharp boundary between the bystander and the mutating population: the bystander occupies lattice sites $ i \leq L/2$, and all other lattice sites $i>L/2$ are occupied by the mutating population (taken to be all red, fast-growing cells initially). We then track the position $x(t)$ of the invasion front over time. We measure the roughness of the front by calculating the variance of the position:\begin{equation}
\langle[ w(t)]^2 \rangle=\langle  [x(t)-\langle x(t)\rangle]^2 \rangle= \langle [x(t)]^2 \rangle - \langle x(t) \rangle^2,
\end{equation}
where we average  over sufficient runs to ensure convergence. 
In the case of a domain wall between just two strains, perhaps with a difference in growth rates, the domain wall performs a biased random walk \cite{RadDK}. Therefore, we may expect that our position $x(t)$ also performs a diffusive motion in time.  The number fluctuations at the boundary introduce a stochasticity to the motion, while the difference in growth rates provides a deterministic bias. So, for a boundary between two strains, we expect the variance $\sigma_w(t)\equiv \sqrt{\langle [w(t)]^2\rangle}$  to satisfy
\begin{equation}
\sigma_w(t)  \approx \sqrt{D_w t},
\end{equation}
with $D_w$ a diffusion coefficient for the domain wall. Indeed, $x(t)$ itself should perform a biased random walk and we may extract the diffusion constant $D_w$ from a time series of the position $x(t)$ performing a time average of the squared displacements of the interface \cite{diffusivities}. We did this for the simulations shown in Fig.~\ref{fig:dw}. We see that when the population is near a mutational meltdown at $\mu \approx \mu^*$ [Fig.~\ref{fig:dw}(a)], the observed diffusivity is much larger than for a population far away from this transition [Fig.~\ref{fig:dw}(b), with $\mu \ll \mu^*$].  However, a proper measurement of $D_w$  requires an ensemble averaging over many simulation runs and also a longer time series. 

We shall see in the following that a more detailed analysis of the boundary motion will show that $x(t)$ actually performs a \textit{super-diffusive} motion near the mutational meltdown $\mu \approx \mu^*$, with displacements satisfying $\sigma_w(t) \propto t^{\nu}$, with $\nu>1/2$.  Super-diffusivity is not uncommon in spatial population dynamics: In a range expansion, for example, the roughness of the expansion front may  contribute to the motion of sectors of strains, introducing super-diffusivity to the sector boundary motion \cite{DRNPNAS}. However, this super-diffusivity depends on the conditions of the growth, and a diffusive motion often serves as a reasonable approximation \cite{Bryan,KorolevBac}.    Moreover, if we are just thinking about populations living in a fixed one-dimensional geometry, then we expect sector boundary motion to be diffusive. We will find diffusive motion of our sector boundaries everywhere in the $(s,b,\mu)$ parameter space, except near the mutational meltdown $\mu \approx \mu^*$ where the sector motion becomes super-diffusive.

\begin{figure*}[htpb!]
\centering
\includegraphics[width=\textwidth]{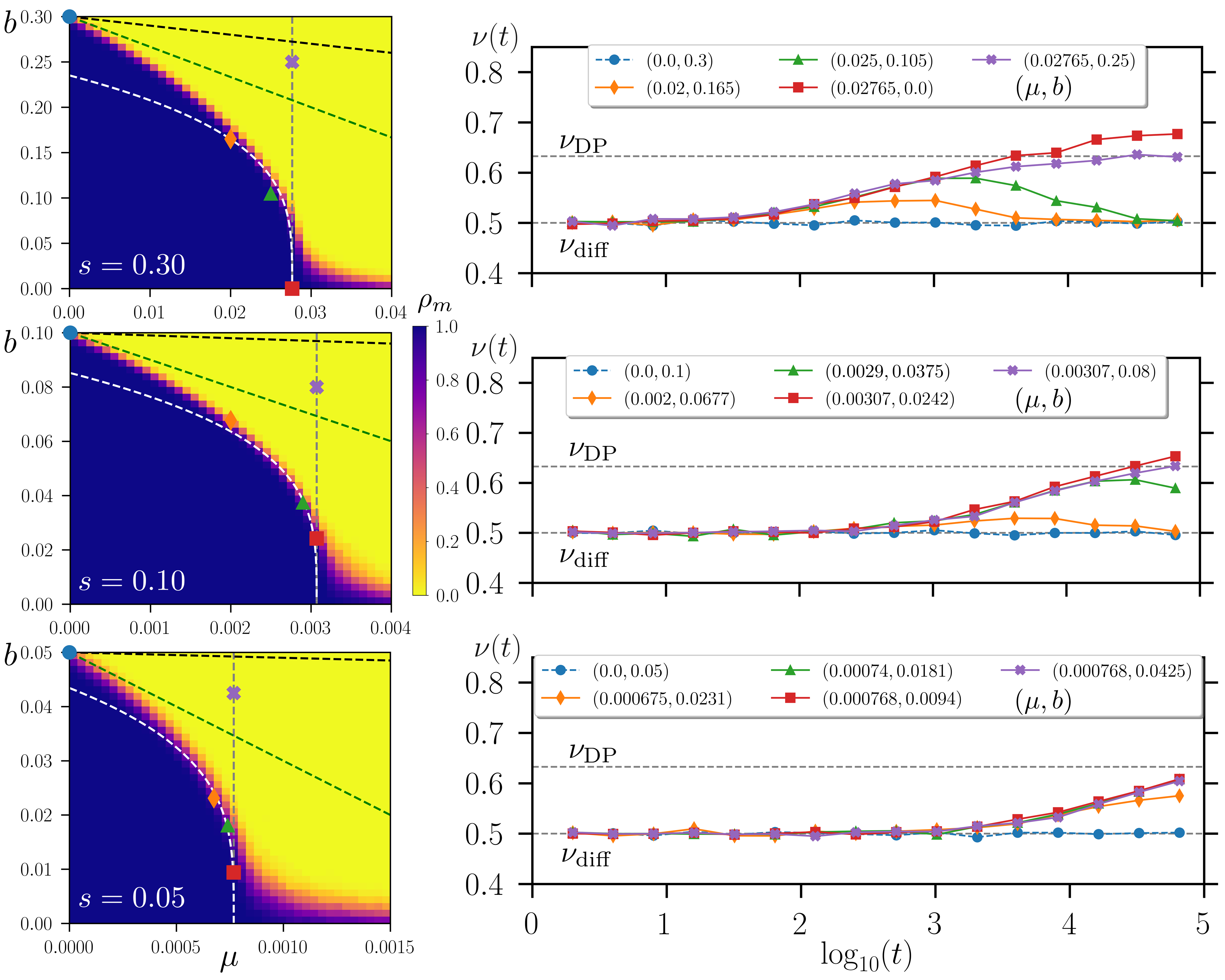}
\caption{For varying values of $s=0.3$ (top row), $0.1$ (middle), and $0.05$ (bottom), we show how, as we move along the critical line starting from $(\mu,b) = (0,s)$ towards the two-species critical point at $(\mu,b) = (\mu^*,0)$, the domain walls separating a bystander population from a mutating one acquire super-diffusive behavior. The phase diagrams on the left illustrate where we calculate the roughness exponent $\nu(t)$ on the right. Two limiting values of the exponent are indicated with dashed lines in each plot: a diffusive $\nu=0.5$ (lower lines) and a super-diffusive, directed percolation value $\nu\approx 0.6326$ (upper lines).   The phase diagrams were created from simulations with $L = 5000$, $t = 10^5$ generations, and  averaged over $256$ runs. The exponent curves on the right were created from simulations with $L=5 \times 10^4$, $t=10^5$ generations, and $400$ runs.}\label{fig:expCompare}
\end{figure*}

Let us now analyze the dynamics in more detail. For a domain wall or invasion front between our mutating, heterogeneous invader population and the homogeneous bystander, the slow- and fast-growing strain patches of the invader will interact differently with the bystander. We can analyze how this impacts the domain wall motion by studying the standard deviation $\sigma_w(t) = \sqrt{\langle [w(t)]^2\rangle}$, averaged over an ensemble of simulation runs.  We   sample our evolved population at times $t=t_i$ and  calculate the effective  exponent associated with the interface width:\begin{equation}
\nu(t = t_i) \equiv \frac{\ln [\sigma_w(t_i)/\sigma_w(t_{i-1})]}{\ln[t_i/t_{i-1}]}, \label{eq:exponent}
\end{equation}
where we choose  $t_i/t_{i-1} \approx 2$. The effective exponent $\nu(t)$ approaches a limiting value at long times. Moreover, any super-diffusive enhancement to the roughness would be seen as a limiting value $\nu > 1/2$.  The exponent is plotted for various values of $(s,b,\mu)$ in Fig.~\ref{fig:expCompare}. We find that there is an enhanced, super-diffusive roughness $(\nu>1/2)$ whenever the mutating population is close to the mutational meltdown (DP) transition at $\mu= \mu^*$ [along the vertical line in the phase diagram in Fig.~\ref{fig:threespeciesPD}(a)]. The enhanced value of $\nu$ near the DP transition may be  understood by considering the limiting case $b=0$. In this case, the bystander population and slow-growing strain within the mutating population will grow at the same rate, so then an initial condition with a single red fast-growing cell in a population of yellow bystander cells will expand as it would in a standard DP process with a single seed initial condition. Hence, the standard deviation $\sigma_w(t)$ scales with the DP\ dynamical critical exponent:  $\sigma_w(t) \sim t^{\nu_{\mathrm{DP}}}$, with $\nu_{\mathrm{DP}} =1/z\approx 0.6326 $ for $d=1+1$ \cite{Hinrichsen}. This exponent is indicated with the upper dashed line in Fig.~\ref{fig:expCompare}.
Introducing a non-zero $b>0$ should not change the situation much; we would only expect a difference in the bias of the domain wall motion.  

\begin{figure}
\centering
\includegraphics[width=0.35\textwidth]{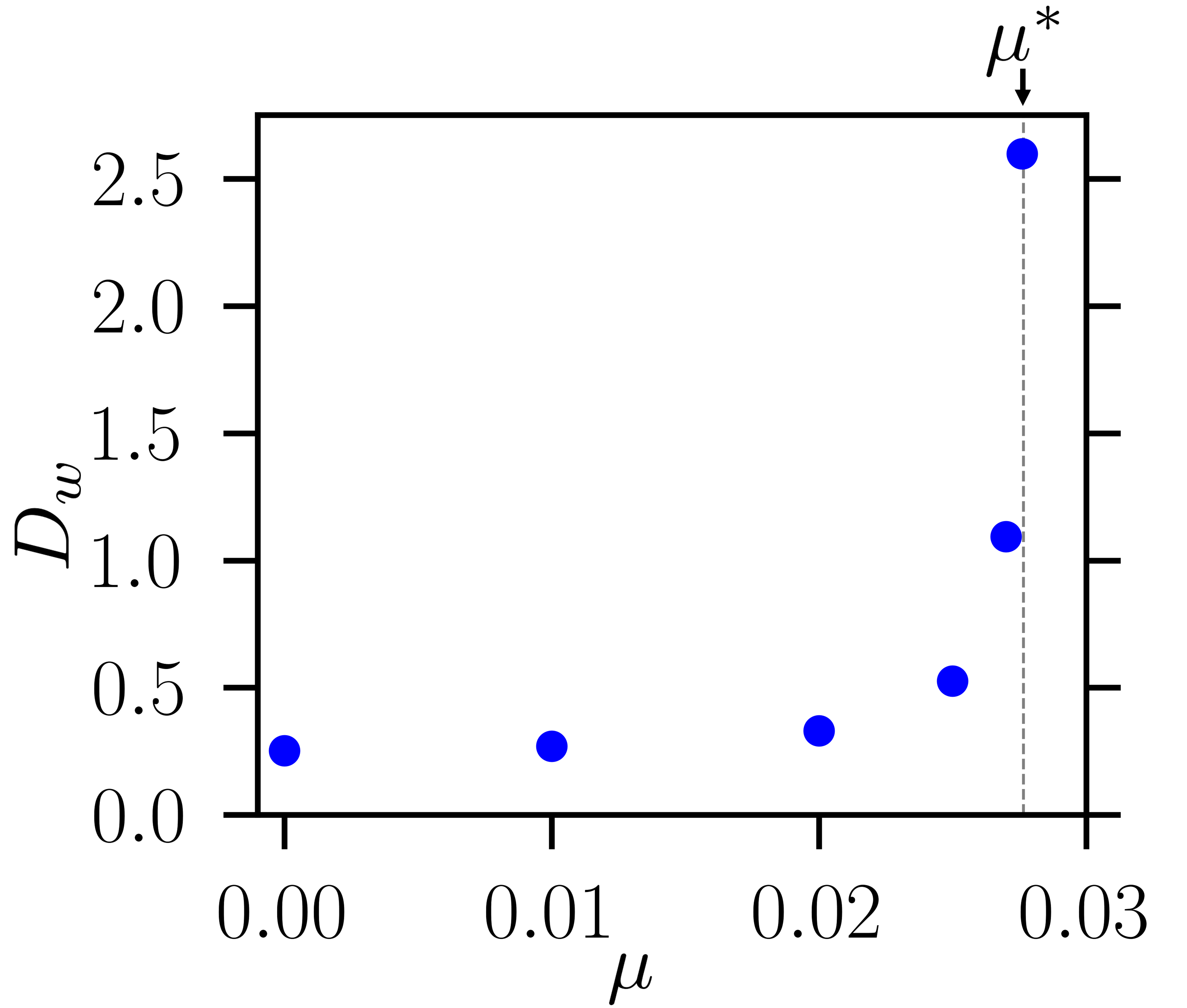}
\caption{\label{fig:diffusivity} The diffusion coefficient, $D_w$, of the boundary between healthy and cancerous populations, measured for various values of $\mu$ approaching $\mu_c$ with $s=0.3$ and $b$ adjusted so that the mutating and bystander populations remain relatively neutral. We see that as $\mu\rightarrow\mu^* \approx 0.02765$, $D_w$ diverges as the domain wall becomes super-diffusive. The coefficients were calculated from simulations with $L=3\times 10^4$, $t \approx 2 \times 10^6$ generations, and averaged over $N = 1280$ runs.} 
\end{figure}

Away from the DP transition, the invasion front  has a diffusive behavior, with $\sigma_w(t) =\sqrt{D_wt}$. The diffusion constant $D_w$ may be measured and serves as a good indicator of the mutational meltdown transition because $D_w$ should diverge as $\mu \rightarrow \mu^*$ for fixed $b$ and $s$. This is illustrated in Fig.~\ref{fig:diffusivity} for $s=0.3$ and values of $b$ along the phase transition boundary.   In this $d=1+1$-dimensional case, the value of $b$, according to our simplified analysis, does not change the wandering behavior of the domain walls as it only serves to change the domain wall \textit{bias}. This hypothesis is consistent with the data shown in Fig.~\ref{fig:expCompare}, where the red squares and purple crosses, despite having very different $b$ values, both exhibit super-diffusive exponents $\nu(t)>1/2$ at long times because both points are near the mutational meltdown transition at $\mu=\mu^*$. We do see small $b$-dependent differences, but these may be due to  the finite time of our simulations. Indeed, the (super-)diffusive scaling of the interface motion only holds at sufficiently long times. It is clear that, especially for the smaller values of $s$ in Fig.~\ref{fig:expCompare}, that the exponent $\nu(t)$ has not yet saturated to its long-time, limiting value within the simulation time. In any case, it is clear that we find super-diffusive motion whenever $\mu$ approaches $\mu^*$ in the various panels of Fig.~\ref{fig:expCompare}.  As we shall see in the next subsection, the situation changes dramatically for the $d=2+1$-dimensional case where the interface between the bystander and mutating populations will behave differently depending on the value of $b$.

\subsection{$d=2+1$-dimensional invasions}

 \begin{figure}[htp]
\centering
\includegraphics[width=0.6\textwidth]{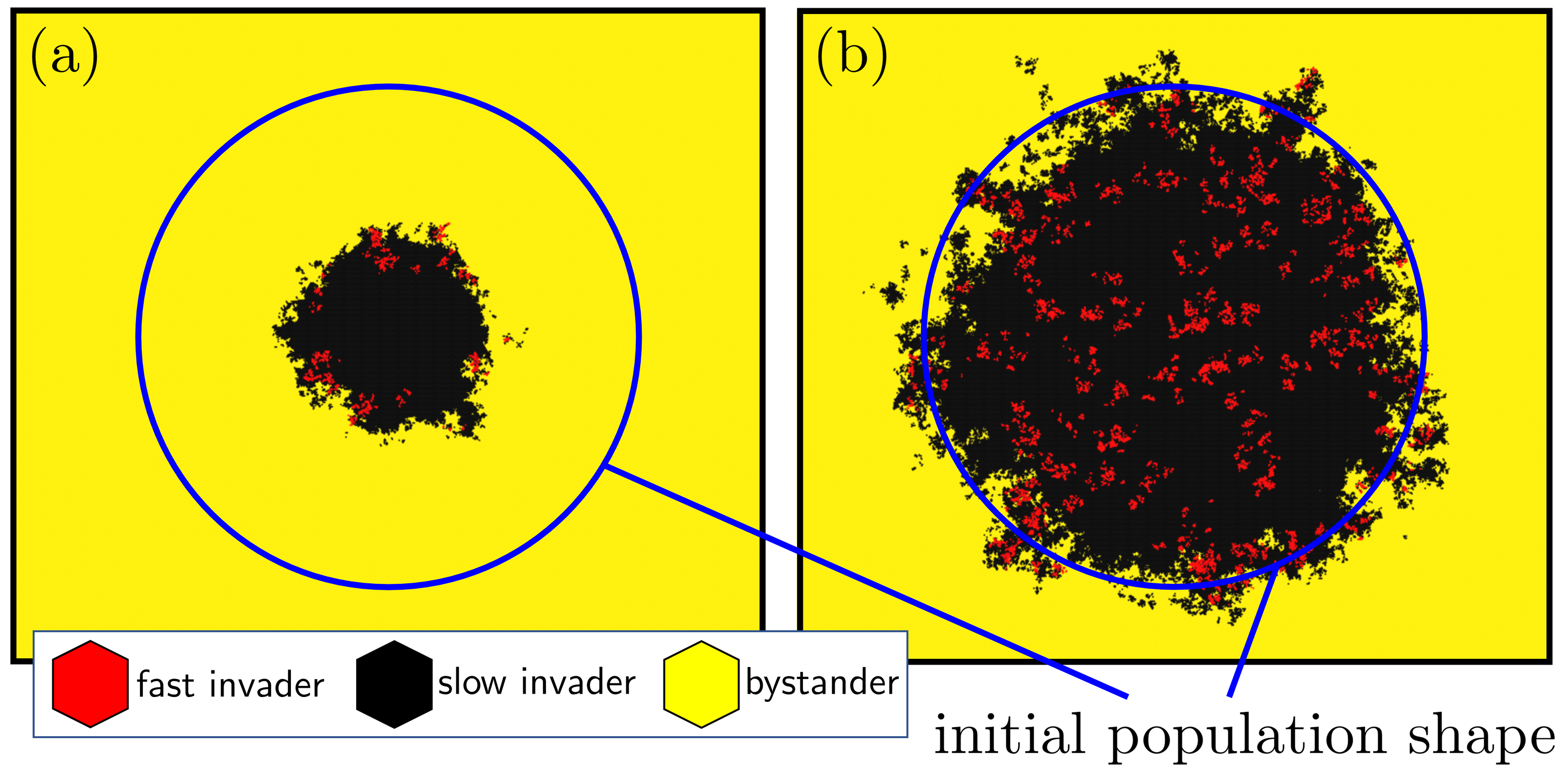}
\caption{Snapshots at $t = 8192$ generations of an initially circular cluster of the fast-growing strain (initial diameter of 400 cells). In (a), the internal parameters $(\mu,s)$ of the cancerous population are set such that there is an average bias of the interface so the growth speed $|v|>0$. In (b), we have $|v|=0$, and thus we are on the critical line of the 3-species phase diagram. In this case, the interface between  populations dissolves, thus our characterization of the interface roughness becomes more complicated in the $d=2+1$-dimensional case than it was for the $d=1+1$-dimensional case, where the boundary was a single point performing a biased (super)diffusive random walk.}\label{fig:disc}
\end{figure}

\begin{figure}[htp]
\centering
\includegraphics[width=0.4\textwidth]{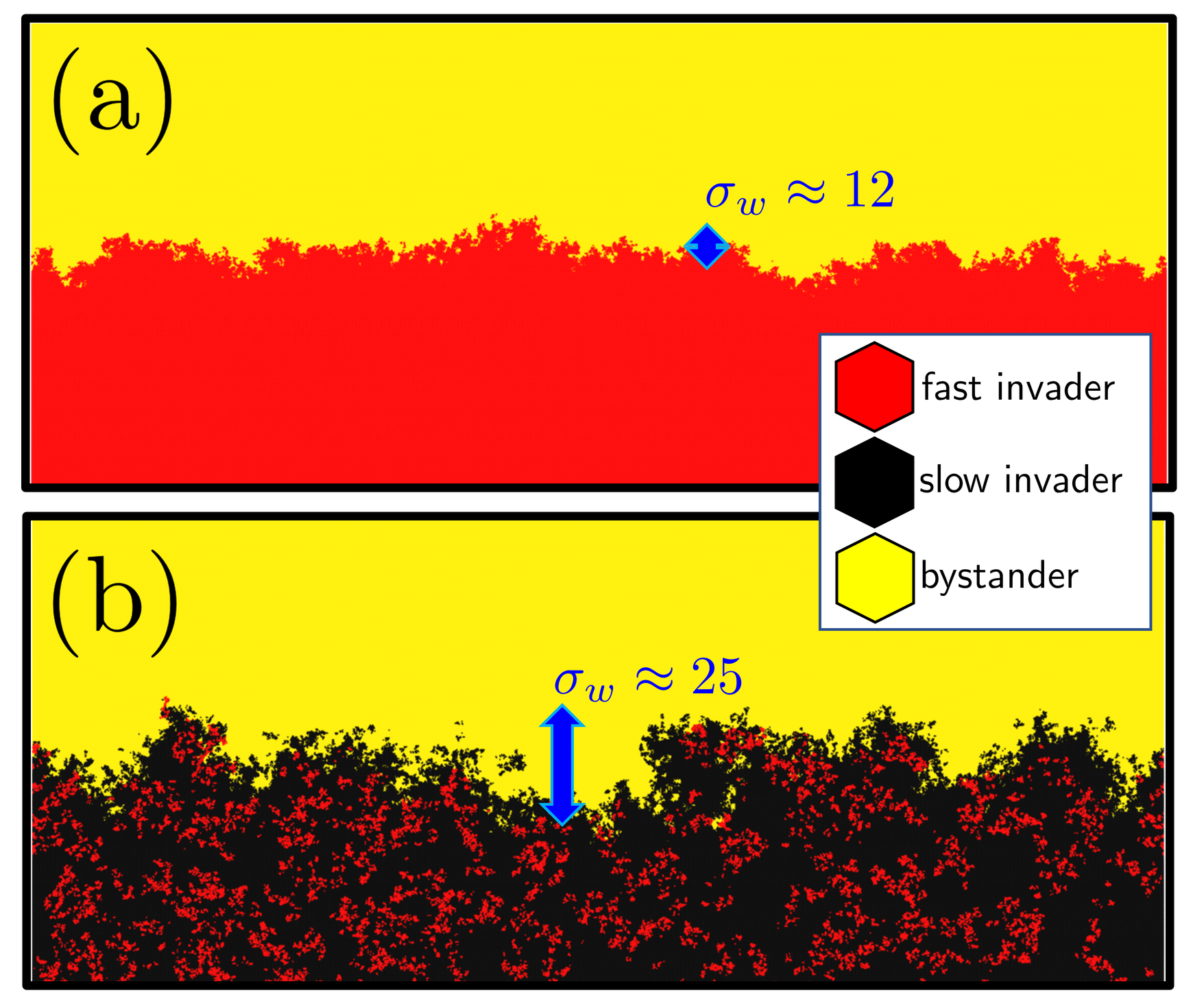}
\caption{A  comparison of the invasion front (in a population with a total of $L^2=100^2$  cells) along the critical line (equal growths for the bystander and invader) at $t\approx 4000$ generations, starting from an initially uniform, flat interface. In  (a), a non-mutating invader is in neutral competition with the bystander  ($\Gamma_f=\Gamma_b=1$ and $\mu=0$) and the interface remains overall stationary and  dissolves over time. In (b), we are just below the mutational meltdown transition with $\mu \approx 0.053$ (for $s=0.15$) for the invader. We set $b=0$ so that the interface is  stationary. We see dissolving of the interface, but there is an enhanced roughness due to the mutational meltdown. The roughness may be quantified directly for these simulation snapshots via the width $\sigma_w$ as defined in Eq.~\eref{eq:width}.  Note that width (given in cell diameters) in (b) is twice that in (a).  }\label{fig:v0_comparison}
\end{figure}

For a two-dimensional population, the invasion front is no longer a simple point, but rather an undulating \emph{line} between the bystander and the mutating red/black population. Moreover, this line can \textit{thicken} as pieces of the invader population pinch off and migrate into the bystander population due to rearrangements induced by the cell division. This dissolution of the front is more prominent when the bystander and the invader have approximately the same growth rates. An example of the complicated frontier shapes are shown in Figs.~\ref{fig:disc}  and \ref{fig:v0_comparison}.  In Fig.~\ref{fig:disc} an initially circular mutating population gets reinvaded by the yellow bystander population in (a) and is approximately neutral with respect to yellow population in (b). We see that in (b) the initially  circular interface dissolves. In (a), the dissolution is less prominent, but still has an effect. This difference in roughness properties between Fig.~\ref{fig:disc}(a) and (b) indicates that the overall growth speed $v$ of the interface between the bystander and the invader will influence the interface roughness.  This is in marked contrast to the $d=1+1$-dimensional case where the average difference in growth rates between the invader and bystander only changes the bias of the (super-)diffusive motion of the interface. Thus, for $d=2+1$, we will have to consider the dynamics at particular values of the average interface speed $v$.

We begin with some qualitative observations of the dynamics. For  relatively neutral populations with an average interface speed $v=0$,  we again expect to see an enhanced roughening of the interface over time when the invader population approaches a mutational meltdown, much like in the $d=1+1$-dimensional case. We can see the enhanced roughening qualitatively in Fig.~\ref{fig:v0_comparison} for this special case where the invader and bystander have the same growth rate [i.e., we are on the phase transition boundary in Fig.~\ref{fig:threespeciesPD}(b) and $v=0$]. In Fig.~\ref{fig:v0_comparison}(a) we have a non-mutating invader and in Fig.~\ref{fig:v0_comparison}(b) we have an invader near a mutational meltdown transition.   The frontier is more undulated in Fig.~\ref{fig:v0_comparison}(b) near the meltdown transition. The increased undulation may be quantified by studying the average interface width $\sigma_w$, which we now describe. 

\begin{figure}[htpb!]
\centering
\includegraphics[width=0.5\textwidth]{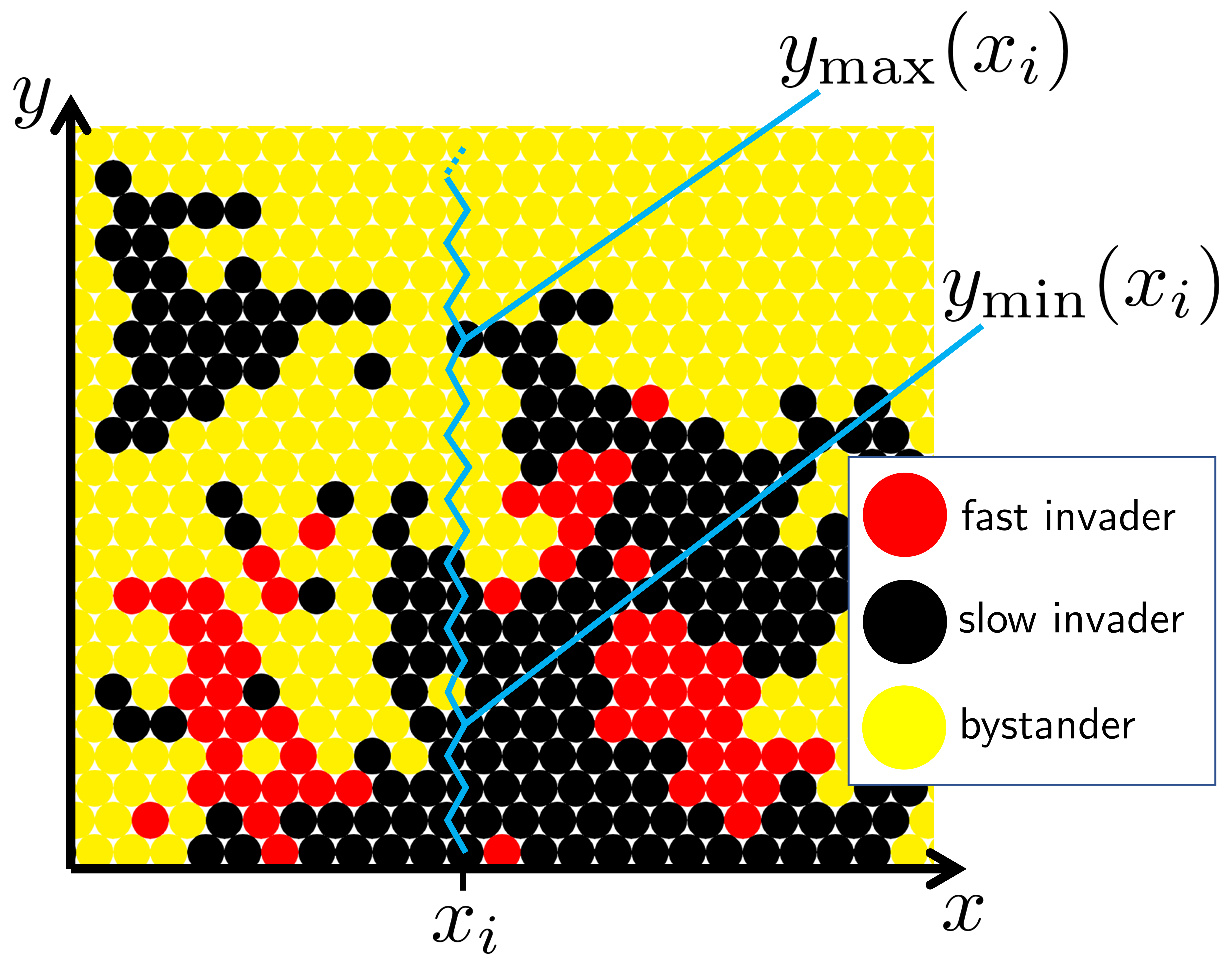}
\caption{Schematic for finding an average location of the interface between the yellow bystander  and red/black mutating populations. The interface runs along the $x$ direction.  We identify  columns  $x_i$ in the hexagonal lattice as shown with the blue zigzagged line. At each column $x_i$, the average  position $\overline{y}$ is calculated by averaging over  all red/black cell locations between the red/black cell which is the furthest into the mutating region [at $y_{\mathrm{min}}(x_i)$] and the black/red cell which is the furthest into the bystander population [at $y_{\mathrm{max}}(x_i)$].  }\label{fig:avg_rules}
\end{figure}

In order to partially mitigate the effects of the ``fuzzing out'' of the interface, we quantify the roughening by looking at the average location of the interface at each time $t$ during the evolution.   To do this, we  set up a coordinate system where we orient a linear population interface along the $x$-direction of our lattice and we let $x_i$ represent the zigzagged columns of our hexagonal lattice along this direction, as shown in Fig.~\ref{fig:avg_rules}. Then, for each column $x_i$, we define the  interface location by averaging over all $N_u$ locations of red/black  cells within a certain range:
\begin{equation}
 \overline{y}(x_i,t)=\frac{1}{N_u}\sum_{y=y_{\mathrm{min}}}^{y_{\mathrm{max}}}y(x_i,t),
 \end{equation} where   $y_{\mathrm{min}} = y_{\mathrm{min}}(x_i,t)$ is  the location of a black/red cell on the lattice at the point $x_i$ (and  time $t$) such that all cells with $y<y_{\mathrm{min}}$ are also black or red. Similarly, $y_{\mathrm{max}}$ is the location of the red/black cell for which all cells with $y>y_{\mathrm{max}}$ are all yellow. The scheme is illustrated in Fig.~\ref{fig:avg_rules}. 

\begin{figure}[htp]
\centering
\includegraphics[width=0.8\textwidth]{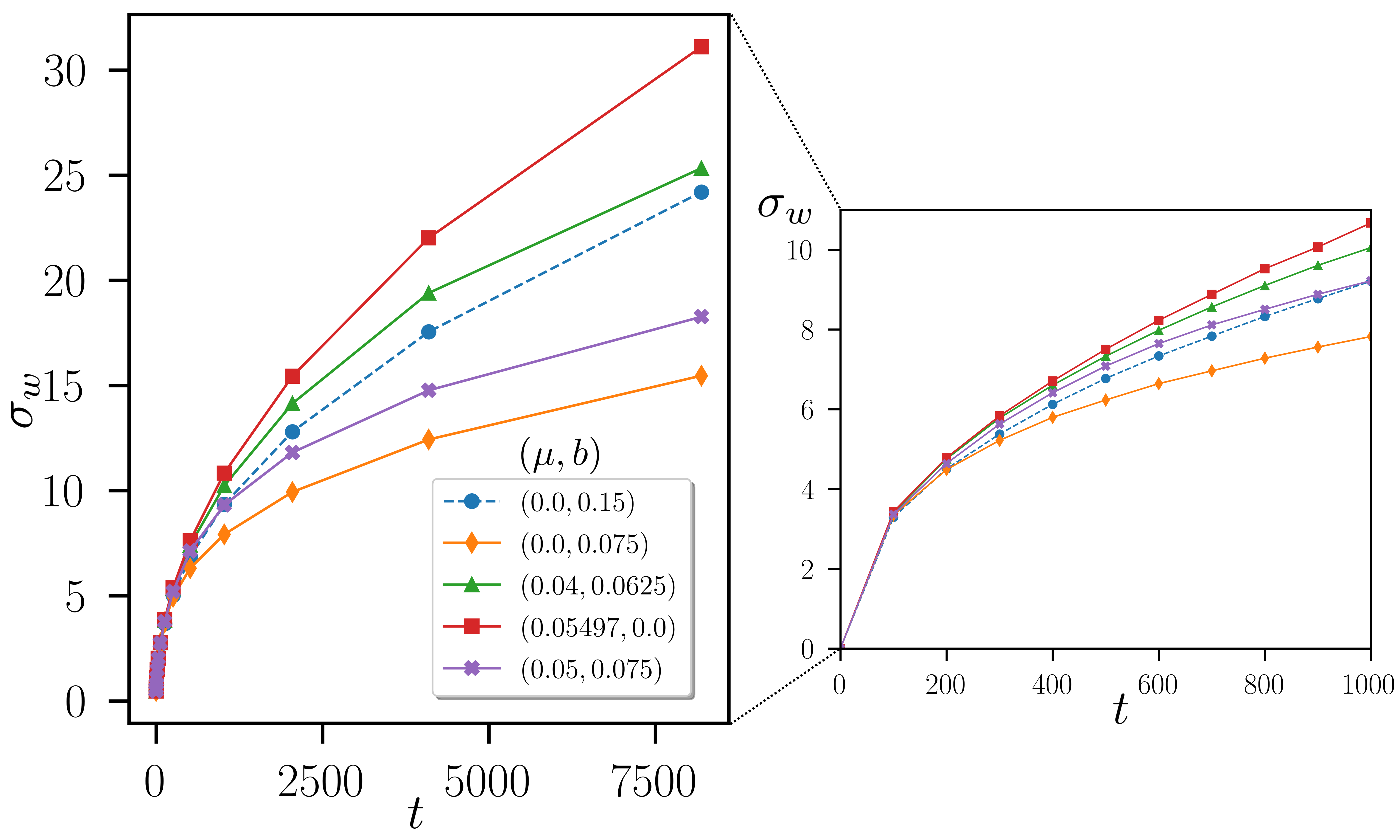}
\caption{The  interface width  $\sigma_w$ [see  Eq.~\eref{eq:width}] in units of cell diameters of a $d=2+1$-dimensional invasion, starting from an initially flat interface between the mutating and bystander population  ($\sigma_w=0$)  $4000$ cells long and with $s=0.15$ for various values of selection parameter $b$  and mutation rate $\mu$.  [Note that it is helpful to consult  the phase diagram in the top panel of Fig.~\ref{fig:3D_phasediagram_critexponent} for identifying the locations of these points in the $(\mu,b)$ plane.]  The interface is evolved for $t=10^4$ generations,  and we average over   $160$ runs.  Lines connect the points to guide the eye. Note that the value of $b$ strongly influences the behavior of $\sigma_w$, as  seen by comparing the red squares and the purple crosses, both of which have the mutating population near meltdown $(\mu\approx\mu^*)$. In general, we find a suppressed roughness when the mutating and bystander populations are not relatively neutral (compare blue dashed line to orange diamonds and purple crosses). Otherwise, for (on average) stationary interfaces, we see the enhanced roughness due to population meltdown (green triangles and red squares). The smaller plot shows the roughness at short times.  \label{fig:w_powerlaw}}
\end{figure}

Using the average location $\overline{y}(x_i,t)$ allows us to define an  interface width $\sigma_w(t)$ by averaging over all  $x_i$ along the interface: 
\begin{equation}\label{eq:width}
\sigma_w(t) =\sqrt{ \left\langle \frac{1}{L}\sum_{x_i = 0}^{L} \left[ \overline{y}(x_i,t) - \overline{\overline{y}}(t)  \right]^2 \right\rangle},
\end{equation}
where 
\begin{equation}\label{eq:l_bar}
\overline{\overline{y}}(t) = \frac{1}{L}\sum_{x_i = 0}^{L} \overline{y}(x_i,t).
\end{equation}
The angular brackets in Eq.~\eref{eq:width} indicate an ensemble average over many population evolutions. However, we may also use  $\sigma_w(t)$ as an indicator of the front roughness for a single snapshot of a population at a particular time, as shown in Fig.~\ref{fig:v0_comparison}. Examples of the calculated $\sigma_w(t)$ (averaged over many simulation runs) for various values of selection parameter $b$ and mutation rate $\mu$ are shown in Fig.~\ref{fig:w_powerlaw}.   For example, in the case where the invader and bystander populations are relatively neutral and there are no mutations, the roughening of the interface illustrated in Fig.~\ref{fig:v0_comparison}(a) is shown with blue circles (connected by a dashed line) in Fig.~\ref{fig:w_powerlaw}. The interface  in Fig.~\ref{fig:v0_comparison}(b) approximately corresponds to the red squares in Fig.~\ref{fig:w_powerlaw}. Note that, as expected, $\sigma_w(t)$ increases significantly faster in time for the latter case compared to the former.

Note that it is possible to define the interface width $\sigma_w$ in other ways, including estimating the interface position using the location $y_{\mathrm{min}}$ or $y_{\mathrm{max}}$ (see Fig.~\ref{fig:avg_rules}). Alternatively, one might use the difference $y_{\mathrm{max}}-y_{\mathrm{min}}$ as a measure of the ``fuzziness'' of the interface, which we might also expect to roughen near a mutational meltdown. We have verified that using other definitions of the interface roughness does not change the long-time scaling properties of the interface roughness or the relative enhancement of the roughness near a mutational meltdown. It would be interesting, however, to more systematically study the consequences of using alternative definitions of the roughness.

  We will now focus our quantitative analysis on the $v=0$ case of a stationary (on average) interface, since it is along the critical line where we find a predictable roughening effect. We will then take a closer look at the cases $|v|>0$ where either the mutating population or the bystander has an overall selective advantage. This introduces complications as the roughening behavior of a moving front is different from a stationary one.
Indeed, whenever $|v|>0$, the invasion becomes a noisy Fisher wave which has its own particular roughening properties. We shall see that a non-zero velocity $v$ will   suppress the interface roughness at long times, but signatures of the roughening due to mutational meltdown persist at shorter times.  

\subsubsection{Voter model coarsening, $v=0$}

Along the 3-species critical surface, where the invader and bystander are relatively neutral, we expect to see an enhancement of the interface roughening as we approach the mutational meltdown transition $\mu\rightarrow\mu^*$ for the invader population [the bottom terminal end of the phase boundary in Fig.~\ref{fig:threespeciesPD}(b)].  To quantify the roughening, we can calculate the effective exponent $\nu(t)$ [see Eq.~\eref{eq:exponent}] from the interface width $\sigma_w(t)$ defined in Eq.~\eref{eq:width}.  Without a bias, we expect that the interface coarsening should be described by voter-model-like dynamics \cite{voterPRL} because the invader and bystander populations  divide into each other without a surface tension.   We generally expect a diffusive behavior $\sigma_w \propto \sqrt{t}$ in this case.

\begin{figure*}[htpb!]
\centering
\includegraphics[width=\textwidth]{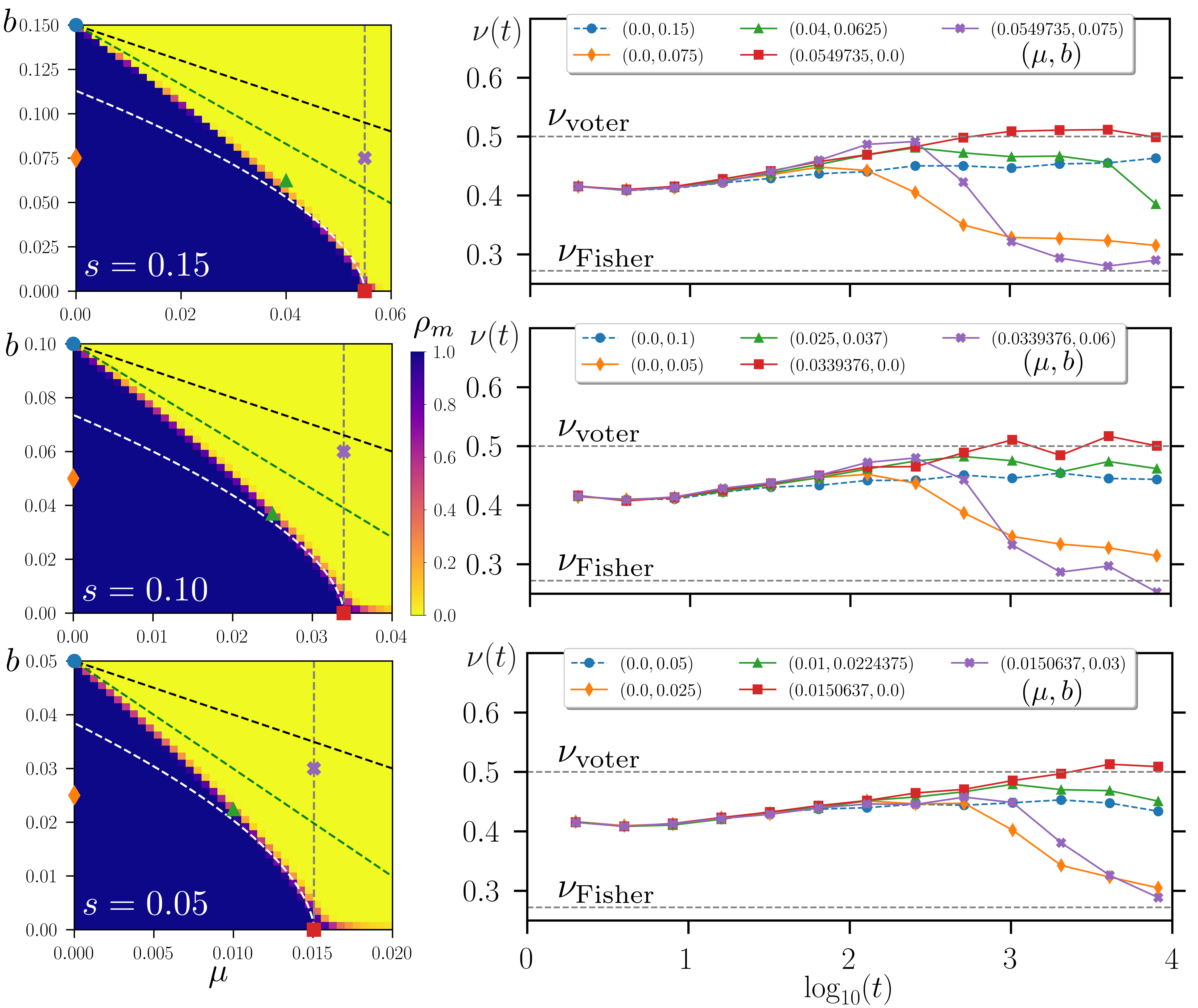}
\caption{Interface roughening exponents $\nu(t)$ are calculated on the right plots for different combinations of $(b,\mu)$ indicated on the phase diagrams on the left, for  varying values of $s=0.3$ (top row), $0.1$ (middle), and $0.05$ (bottom).  As we move along the critical line (blue circles, green triangles, and red squares),    we show the  enhancement of the boundary roughness   [from $\sigma_w \propto t^{0.4}$ to $\sigma_w \propto  t^{0.5}$].  Away from the critical line (purple crosses and orange diamonds), we see the effects of Fisher wave dynamics. Here either the mutating population (orange diamonds) or the bystander (purple crosses) has a selective advantage, and the moving interface has a suppressed roughness at long times, approaching $\sigma_w(t) \propto t^{0.272}$ (bottom dashed lines in plots on the right), consistent with previous Fisher wave simulation results \cite{benavraham}. The phase diagrams have the same simulation parameters as in Fig.~\ref{fig:PDcompare}. The exponent curves on the right use interfaces that are initially $4000$ cells long, and we average over 160 runs. }\label{fig:3D_phasediagram_critexponent}
\end{figure*}

In Fig.~\ref{fig:3D_phasediagram_critexponent} we see an enhanced roughening as $\mu\rightarrow\mu^*$ as we move along the phase transition boundary ($v=0$): The limiting value $\nu$ of the exponent increases as we move along the phase transition line towards the mutational meltdown at $\mu = \mu^*$. Interestingly, near mutational meltdown, the width $\sigma_w$ seems to grow approximately diffusively with  $\sigma_w \propto t^{0.5}$ (red squares in Fig.~\ref{fig:3D_phasediagram_critexponent}), whereas the non-mutating case $\mu=0$ coarsens according to the power law $\sigma_w \propto  t^{0.4}$ (blue circles in Fig.~\ref{fig:3D_phasediagram_critexponent}). We might have expected larger values for these exponents as the non-mutating case should be closest to the voter model dynamics where interfaces dissolve diffusively, similarly to the dynamics of $\sigma_w$ in the $d=1+1$ case away from the meltdown transition \cite{voterPRL}.  However, generalizations of the voter model can yield different results for interface coarsening and determining the value of the exponent $\nu$ can be subtle \cite{votercoarsening}. Another possibility is that $\nu$ is suppressed due to our particular choice of lattice update rules. It would be interesting to study the behavior with simulations with overlapping generations (independently dividing cells).

Although the behavior for $d=2+1$ is different from the $d=1+1$ case where the domain wall roughening was clearly super-diffusive near mutational meltdown and diffusive away from it (see Fig.~\ref{fig:expCompare}), we also find here that the mutational meltdown enhances the interface undulations by modifying the exponent $\nu$ associated with the interface width $\sigma_w \propto t^{\nu}$, increasing $\nu$ from a value of approximately 0.4 to 0.5.    A more dramatic difference is found when we move away from the $v=0$ critical line and have either the mutating invader population or the bystander grow with an overall selective advantage. We then find a moving Fisher wave with a suppressed exponent $\nu$, as we will see in the next section.

\subsubsection{Fisher wave roughening, $|v| > 0$}

The comparison between $|v|>0$ and $v \approx 0$ dynamics can be seen prominently if we consider an initially disc-like population of the invader strain. Then, any non-zero velocity will either shrink or grow the initial disc. An example of an $v<0$  evolution is shown in Fig.~\ref{fig:disc}(a) where the bystander strain reinvades the invader, which eventually dies out. Conversely, when $v \approx0$, we can see in Fig.~\ref{fig:disc}(b) that the boundary between the invader and bystander   gradually dissolves. This illustrates the key feature that makes $|v|>0$ different from the critical line: one of the populations (either the mutating population or the bystander) becomes \textit{unstable} and will deterministically shrink in the presence of the other population.

Let us consider first the simplest case when $\mu=0$ and we have an interface between a (non-mutating) fast-growing red strain and the yellow bystander. The orange diamond point data in Fig.~\ref{fig:3D_phasediagram_critexponent} show what happens in this case.  The interface behaves as a noisy Fisher-Kolmogorov-Petrovsky-Piskunov wave \cite{fisher,KPPeq} describing the invasion of the bystander.  Without fluctuations (in the mean field limit), these waves admit stationary shapes and we   have no roughening over time. However, fluctuations prevent the formation of stationary wave fronts for the $d=1+1$ and $d=2+1$-dimensional cases. For $d=2+1$,  previous simulations \cite{benavraham} show that the interface width is expected to grow as $t^{\nu}$ with $\nu \approx 0.272$. This coarsening is consistent with our results for $\sigma_w$,  as the orange diamond data points  in the right panels of Fig.~\ref{fig:3D_phasediagram_critexponent} approach the $\nu \approx 0.272$ limiting value at long times, indicated by the lower dashed line. The time until convergence, however, is quite long as the effective exponent $\nu(t)$ continues to decrease over the course of the entire simulation run time.

The   case of a non-mutating invader is interesting for $d=2+1$ because we would naively expect our system to fall into the KPZ universality class. The average interface position $\overline{y}(x_t,t)$ could be interpreted as a kind of ``height function'' and the interface width $\sigma_w$ should scale like $\sigma_w \propto t^{1/3}$ at early times, consistent with $d=1+1$-dimensional KPZ dynamics. A broad class of systems fall into this universality class (see \cite{KPZreview} for a review) as the KPZ equation includes the most relevant nonlinearity associated with lateral growth. However, we see here that the behavior is more subtle as the fuzzing out of the interface will contribute to the measured roughness. This complication in measuring the interface roughness was discussed and analyzed in previous work \cite{MoroFisher}.  Our focus here, however, is not the particular  exponent associated with the roughening but rather the effects of adding mutations.
We will see that adding mutations does   enhance the roughness, but only at short/intermediate times while the fast-growing, mutating strain maintains a significant fraction within the population.

Consider the portion   of the phase diagram where the bystander can reinvade the mutating population due to fitness loss at a non-zero mutation rate $\mu$ (purple crosses in Fig.~\ref{fig:3D_phasediagram_critexponent}). Here, the evolution of our system begins at first as biased competition between two species  (between \emph{fast-growing} and \emph{bystander} species) but as the \emph{fast-growing} cells mutate and die off, the bystander population begins to reinvade the  \emph{slow-growing} species, and eventually we should find a Fisher wave of the bystander invading the less fit, mutating population. On the right side of Fig.~\ref{fig:3D_phasediagram_critexponent} we see that the  exponent $\nu(t)$ for the purple crosses at first is enhanced as we would expect near mutational meltdown ($\mu \approx \mu^*$). At later times, however, once a Fisher wave is established, the exponent eventually dips down and is consistent with a Fisher-wave-like coarsening \cite{benavraham}. One can track this especially easily in the $s=0.15$ case (top row of Fig.~\ref{fig:3D_phasediagram_critexponent})
where we see that the purple cross data points follow the critical roughening points (red squares) and then transition to a slower roughening more consistent with a regular Fisher wave (orange diamonds).
The evolution of $\sigma_w(t)$ for this case is also shown in Fig.~\ref{fig:w_powerlaw}.
 One sees here that at times $t<1000$ (smaller plot), the purple cross data points have a larger width $\sigma_w(t)$ due to the mutational meltdown dynamics, but $\sigma_w(t)$ then crosses over to smaller values for longer times when the Fisher wave behavior dominates.

\section{Conclusion \label{sec:conclusion}}

We have now analyzed a simple model of invasion of a stable, homogeneous population by a    population acquiring deleterious mutations at a rate $\mu$. We examined this invasion in both one- and two-dimensions as a function of the mutation rate $\mu$, the selective advantage $s$ of the fast-growing strain within the mutating population, and the selective advantage $b$ of the bystander  population. We have shown that the effectively small local population sizes (compared to a well-mixed population) suppress the probability that the invasion succeeds. This suppression can be understood by analyzing the motion of the boundary between the mutating population and the bystander population it is invading. We find a reasonable estimate of the phase transition position in the $(\mu,b,s)$ phase space,  as shown in Figs.~\ref{fig:threespeciesPD},\ref{fig:expCompare}, and \ref{fig:3D_phasediagram_critexponent}. Our model assumed that cell motility within our population is suppressed, with the only cell rearrangements occurring due to cell division and local competition for space. It would be interesting to consider the effects of a spatial motility as it has been demonstrated that some of the expected features of spatial dynamics, such as spatial heterogeneity and local fixation of strains is partially mitigated by increased cell motility \cite{WaclawCancer}.

Next, we considered the properties of the invasion front and showed that this front undulates more when the mutating population is near the  meltdown transition at which it loses the fittest strain. For $d=1+1$ dimensions, this transition is well-characterized by the directed percolation universality class, and we used properties of this class to understand the enhancement of the roughening. In the future, it would be interesting to compare our results to experiments. One possibility is to use microbial populations such as bacteria or yeast where one may design strains with varying $(\mu,b,s)$. Another possibility would be to examine such invasions in cancers. For instance, it would be interesting to monitor the edges of a tumor over time as it either grows or shrinks. We predict that if the tumor begins losing fitness due to accumulated deleterious mutations (during  treatment, for example), then we should be able to observe this transition to ``mutational meltdown'' as a roughening of the tumor edges.

For $d=2+1$ dimensions, we find a  range of behaviors for the roughening interface. When the speed of the invasion front approaches zero, the interface roughens more significantly due voter-model-like coarsening. We also find an enhancement of the roughening exponent as the mutating population approaches meltdown. On the other hand, when either the mutating or the bystander population has a selective advantage and the population interface develops an overall velocity, the roughening is  suppressed, and we find roughening exponents consistent with those observed for noisy Fisher waves at long times. Therefore, the long-time behavior of the population interface roughness serves as an indicator of whether or not a selective sweep is occurring within the population: Moving population fronts will be smoother than stationary ones in which the invader and bystander populations are relatively neutral.

At intermediate times, we see signatures of the meltdown as the interface roughens more rapidly when the mutating population is near the meltdown transition, even in the case when there is an overall bias to the interface motion (see smaller plot of Fig.~\ref{fig:w_powerlaw}). Also, we focused here on just one aspect of the roughening, namely the early time behavior $\sigma_w \propto t^{\nu}$. For long times $t$, the interface undulation size will eventually saturate due to the finite system size $L$, and we might expect a general scaling form $\sigma_w = t^{\nu}f(t/L^{\beta})$, with $f(x)$ a scaling function and $\beta$ a new critical exponent. The scaling properties of this saturation should also depend on the proximity to the mutational meltdown transition.  It would also be interesting to consider a $d=3+1$-dimensional evolution such as the invasion of surrounding tissue by a compact cluster of  cancerous cells. In this case, the invasion front would be an entire \textit{surface} which could also pinch off and coarsen. Previous simulations of the noisy Fisher wave dynamics suggest that the situation in this case is similar to the  $d=2+1$ case considered here \cite{benavraham}. We would again expect to find some enhancement of the interface roughening when a mutating invader is near a mutational meltdown transition.

Interestingly,   increased roughening is typically an indicator of more malignant cancerous growths, and the roughness of tumor edges has been a useful prognostic indicator in a wide variety of cancers \cite{roughnessreview}.  Also, in general, increased heterogeneity results in a worse clinical prognosis \cite{prognosis}. While our results point to the possibility of an opposite correlation, our model does not take into account tumor vasculature or cancer cell motility. Conversely, most of the clinical studies focus on more mature tumors which have developed a vasculature. Hence, we expect our model to be relevant for early, small avascular tumors or regions of larger tumors lacking vasculature. These small tumors are not easily detected as they are typically just a few millimeters in size. Nevertheless, small spheroidal avascular tumors are  good \textit{in vitro} models for early cancer growth \cite{spheroid}. It would thus be interesting to study the edges of such cultured tumors under a large mutational load. We may also verify some of our results in microbial range expansions (e.g., in yeast cell colonies grown on Petri dishes) where there is little cell motility. A promising experimental realization of a $d=2+1$-dimensional expansion may be a growing cylindrical ``pillar'' of yeast cells, as realized in Ref.~\cite{pillar}.

\ack

M.O.L. thanks B. Weinstein for helpful discussions. Computational support was provided by the University of Tennessee and Oak Ridge National Laboratory's National Institute for Computational Sciences. M.O.L. acknowledges partial funding from the Neutron Sciences Directorate (Oak Ridge National Laboratory), sponsored by the US Department of Energy, Office of Basic Energy Sciences.

\section*{References}

\bibliographystyle{iopart-num-long}
\bibliography{mainref}
\end{document}